\PassOptionsToPackage{unicode}{hyperref}
\PassOptionsToPackage{hyphens}{url}
\PassOptionsToPackage{dvipsnames,svgnames,x11names}{xcolor}
\documentclass[
  a4paper,
  DIV=11,
  numbers=noendperiod,
  french]{scrartcl}

\usepackage{amsmath,amssymb}
\usepackage{iftex}
\ifPDFTeX
  \usepackage[T1]{fontenc}
  \usepackage[utf8]{inputenc}
  \usepackage{textcomp} % provide euro and other symbols
\else % if luatex or xetex
  \usepackage{unicode-math}
  \defaultfontfeatures{Scale=MatchLowercase}
  \defaultfontfeatures[\rmfamily]{Ligatures=TeX,Scale=1}
\fi
\usepackage{lmodern}
\ifPDFTeX\else  
\fi
\IfFileExists{upquote.sty}{\usepackage{upquote}}{}
\IfFileExists{microtype.sty}{% use microtype if available
  \usepackage[]{microtype}
  \UseMicrotypeSet[protrusion]{basicmath} % disable protrusion for tt fonts
}{}
\makeatletter
\@ifundefined{KOMAClassName}{% if non-KOMA class
  \IfFileExists{parskip.sty}{%
    \usepackage{parskip}
  }{% else
    \setlength{\parindent}{0pt}
    \setlength{\parskip}{6pt plus 2pt minus 1pt}}
}{% if KOMA class
  \KOMAoptions{parskip=half}}
\makeatother
\usepackage{xcolor}
\usepackage[top=2cm,bottom=2cm,left = 2.5cm,right = 2.5cm]{geometry}
\ifx\paragraph\undefined\else
  \let\oldparagraph\paragraph
  \renewcommand{\paragraph}[1]{\oldparagraph{#1}\mbox{}}
\fi
\ifx\subparagraph\undefined\else
  \let\oldsubparagraph\subparagraph
  \renewcommand{\subparagraph}[1]{\oldsubparagraph{#1}\mbox{}}
\fi

\providecommand{\tightlist}{%
  \setlength{\itemsep}{0pt}\setlength{\parskip}{0pt}}\usepackage{longtable,booktabs,array}
\usepackage{calc} % for calculating minipage widths
\usepackage{etoolbox}
\makeatletter
\patchcmd\longtable{\par}{\if@noskipsec\mbox{}\fi\par}{}{}
\makeatother
\IfFileExists{footnotehyper.sty}{\usepackage{footnotehyper}}{\usepackage{footnote}}
\makesavenoteenv{longtable}
\usepackage{graphicx}
\makeatletter
\def\maxwidth{\ifdim\Gin@nat@width>\linewidth\linewidth\else\Gin@nat@width\fi}
\def\maxheight{\ifdim\Gin@nat@height>\textheight\textheight\else\Gin@nat@height\fi}
\makeatother
\setkeys{Gin}{width=\maxwidth,height=\maxheight,keepaspectratio}
\makeatletter
\def\fps@figure{htbp}
\makeatother

\DeclareMathOperator*{\argmin}{argmin}
\DeclareMathOperator*{\argmax}{argmax}
\DeclareMathOperator*{\med}{med}
\DeclareMathOperator{\Cov}{Cov}
\DeclareMathOperator{\tr}{tr}

\newcommand\1{{\mathds 1}}

\newcommand{\transp}[1]{{}^t\!#1}

\newcommand{\E}[1]{\mathbb{E}\left[#1\right]}
\newcommand{\V}[1]{\mathbb{V}\left[#1\right]}
\newcommand{\cov}[2]{\Cov\left(#1,#2\right)}

\usepackage{mathrsfs, dsfont}
\usepackage[style=authoryear,uniquename=false, uniquelist=false]{biblatex}
\DefineBibliographyStrings{french}{andothers={et\addabbrvspace alii}}
\KOMAoption{captions}{tableheading,figureheading}
\makeatletter
\@ifpackageloaded{caption}{}{\usepackage{caption}}
\AtBeginDocument{%
\ifdefined\contentsname
  \renewcommand*\contentsname{Table des matières}
\else
  \newcommand\contentsname{Table des matières}
\fi
\ifdefined\listfigurename
  \renewcommand*\listfigurename{Liste des Figures}
\else
  \newcommand\listfigurename{Liste des Figures}
\fi
\ifdefined\listtablename
  \renewcommand*\listtablename{Liste des Tables}
\else
  \newcommand\listtablename{Liste des Tables}
\fi
\ifdefined\figurename
  \renewcommand*\figurename{Figure}
\else
  \newcommand\figurename{Figure}
\fi
\ifdefined\tablename
  \renewcommand*\tablename{Table}
\else
  \newcommand\tablename{Table}
\fi
}
\@ifpackageloaded{float}{}{\usepackage{float}}
\floatstyle{ruled}
\@ifundefined{c@chapter}{\newfloat{codelisting}{h}{lop}}{\newfloat{codelisting}{h}{lop}[chapter]}
\floatname{codelisting}{Listing}

\usepackage{amsthm}
\theoremstyle{remark}
\AtBeginDocument{}
\newtheorem*{remark}{Remarque}

\makeatother
\makeatletter
\@ifpackageloaded{caption}{}{\usepackage{caption}}
\@ifpackageloaded{subcaption}{}{\usepackage{subcaption}}
\makeatother
\ifLuaTeX
\usepackage[bidi=basic]{babel}
\else
\usepackage[bidi=default]{babel}
\fi
\babelprovide[main,import]{french}
\def\languageshorthands#1{}
\ifLuaTeX
  \usepackage{selnolig}  % disable illegal ligatures
\fi
\usepackage[]{biblatex}
\usepackage{bookmark}

\IfFileExists{xurl.sty}{\usepackage{xurl}}{} % add URL line breaks if available
\hypersetup{
  pdftitle={Estimation de la tendance-cycle avec des méthodes robustes aux points atypiques},
  pdfauthor={Alain Quartier-la-Tente},
  pdflang={fr},
  colorlinks=true,
  linkcolor={blue},
  filecolor={Maroon},
  citecolor={Blue},
  urlcolor={Blue},
  pdfcreator={LaTeX via pandoc}}

\title{Estimation de la tendance-cycle avec des méthodes robustes aux
points atypiques}
\author{Alain Quartier-la-Tente}
\date{}

\begin{document}
\maketitle

\renewcommand{\thepage}{\roman{page}}

\subsubsection*{Résumé}\label{ruxe9sumuxe9}
\addcontentsline{toc}{subsubsection}{Résumé}

Les séries désaisonnalisées sont usuellement utilisées pour l'analyse de
la conjoncture et des points de retournement. Lorsque l'irrégulier est
trop élevé, il est préférable de lisser la série afin d'analyser
directement la composante tendance-cycle. Cette étude s'intéresse à
l'estimation en temps réel de cette composante autour de chocs et de
points de retournement. Les moyennes mobiles linéaires classiquement
utilisées pour l'estimation de la tendance-cycle, sensibles à la
présence de points atypiques, sont comparées à des méthodes
non-linéaires robustes. Nous proposons également une méthodologie pour
étendre les moyennes mobiles de Henderson et de Musgrave afin de prendre
en compte des informations extérieurs et ainsi construire des moyennes
mobiles robustes à la présence de certains chocs. Nous décrivons comment
estimer des intervalles de confiance pour les estimations issues de
moyennes mobiles, ce qui permet de valider l'utilisation de ces nouvelle
moyennes mobiles. En comparant les méthodes sur des séries simulées et
réelles, nous montrons que : construire des moyennes mobiles robustes
permet de réduire les révisions et de mieux modéliser les points de
retournement autour de chocs, sans dégrader les estimations lorsqu'aucun
choc n'est observé ; les méthodes non-linéaires robustes ne permettent
pas d'extraire une composante tendance-cycle satisfaisante pour
l'analyse conjoncturelle, avec parfois des révisions importantes.

Cette étude est entièrement reproductible et tous les codes utilisés
sont disponibles sous \url{https://github.com/AQLT/robustMA}.

Mots clés : séries temporelles, tendance-cycle, désaisonnalisation,
points de retournement.

\subsubsection*{Abstract}\label{abstract}
\addcontentsline{toc}{subsubsection}{Abstract}

Seasonally adjusted series are usually used to analyse the business
cycle and turning points. When the irregular is too high, it is
preferable to smooth the series in order to analyse the trend-cycle
component directly. This study focuses on the real-time estimation of
the trend-cycle component around shocks and turning points. The linear
moving averages classically used for estimating the trend-cycle, which
are sensitive to the presence of atypical points, are compared with
robust non-linear methods. We also propose a methodology for extending
the Henderson and Musgrave moving averages to take account of external
information and thus construct moving averages that are robust to the
presence of certain shocks. We describe how to estimate confidence
intervals for estimates derived from moving averages, thereby validating
the use of these new moving averages. By comparing the methods on
simulated and real series, we show that: building robust moving averages
makes it possible to reduce revisions and better model turning points
around shocks, without degrading the estimates when no shock is
observed; robust non-linear methods do not make it possible to extract a
trend-cycle component that is satisfactory for economic analysis, with
sometimes significant revisions.

This study is fully reproducible and all the codes used are available
under \url{https://github.com/AQLT/robustMA}.

Keywords: time series, trend-cycle, seasonal adjustment, turning points.

JEL Classification: C22, E32.

\newpage
\renewcommand*\contentsname{Table des matières}
{
\hypersetup{linkcolor=}
\setcounter{tocdepth}{2}
\tableofcontents
}
\newpage
\pagenumbering{arabic}

\section{Introduction}\label{introduction}

L'analyse du cycle économique, et en particulier la détection rapide des
points de retournement, est un sujet de première importance dans
l'analyse de la conjoncture économique. Les séries chronologiques se
décomposent en trois composantes : les effets saisonniers et les effets
de calendrier, la tendance-cycle (permettant d'analyser les points de
retournement) et l'irrégulier (le bruit lié à des erreurs
d'échantillonnage, chocs économiques, etc.). Pour l'analyse
conjoncturelle, les indicateurs économiques sont généralement uniquement
corrigés des variations saisonnières et des jours ouvrables, laissant
ainsi l'effet combiné de la tendance-cycle et de l'irrégulier.
Toutefois, lorsque l'irrégulier est trop élevé, il peut être nécessaire
d'effectuer un lissage supplémentaire afin d'analyser directement la
composante tendance-cycle. Par construction, les méthodes d'extraction
de tendance-cycle sont étroitement liées aux méthodes de
désaisonnalisation. En effet, afin d'estimer la composante saisonnière,
les algorithmes de désaisonnalisation estiment préalablement une
composante tendance-cycle. Ainsi, même si les méthodes d'extraction de
tendance-cycle sont généralement appliquées sur des séries corrigées des
variations saisonnières, l'estimation de ces séries dépend également
également des méthodes d'estimation de la tendance-cycle.

Les moyennes mobiles, ou les filtres linéaires, sont omniprésents dans
les méthodes d'extraction du cycle économique et d'ajustement
saisonnier. Par exemple, la méthode de désaisonnalisation X-11
\autocite{ladiray2011seasonal}, utilisée dans le logiciel
X-13ARIMA-SEATS \autocite{x13}, utilise des moyennes mobiles pour
estimer les principales composantes d'une série chronologique. Au centre
de la série, des filtres symétriques sont appliqués (utilisation
d'autant de points dans le passé et dans le futur et le même poids est
associé aux observations passées et futures). Pour l'extraction de la
tendance-cycle, le filtre symétrique le plus connu est celui de
\textcite{henderson1916note}, notamment utilisé dans l'algorithme de
désaisonnalisation X-13ARIMA-SEATS. En revanche, pour les estimations en
temps réel, en raison de l'absence d'observations futures, toutes ces
méthodes doivent s'appuyer sur des filtres asymétriques pour estimer les
points les plus récents. Par exemple, pour l'extraction de la
tendance-cycle tendances-cycles, X-11 utilise le filtre symétrique de
Henderson et les filtres asymétriques de \textcite{musgrave1964set} sur
une série étendue utilisant un modèle ARIMA. Comme les valeurs prédites
sont des combinaisons linéaires des valeurs passées, cela revient à
appliquer des moyennes mobiles asymétriques à la fin de la série.

Ces moyennes mobiles, comme tout opérateur linéaire, sont sensibles à la
présence de points atypiques. L'application directe des méthodes peut
donc conduire à des estimations biaisées, du fait de leur présence.
alors que les méthodes de désaisonnalisation (comme X-13ARIMA-SEATS) ont
un module de correction des points atypiques.

L'objectif de cette étude est d'étudier et de comparer différentes
approches permettant l'extraction de tendance-cycle en temps-réel. Nous
décrivons également construire des moyennes mobiles linéaires, associées
aux moyennes mobiles de Henderson et de Musgrave, qui prennent en compte
des informations extérieures, permettant notamment de modéliser des
chocs et donc de les rendre robustes à ces derniers. La modélisation
retenue peut être validée par la construction d'intervalles de
confiance. Ces moyennes mobiles sont également comparées à des méthodes
non-linéaires robustes.

Après une description des méthodes utilisées pour extraire la
tendance-cycle (section~\ref{sec-methodo}), elles sont comparées sur des
séries simulées et réelles autour de chocs et de points de retournement
(section~\ref{sec-results}). Nous montrons alors que les moyennes
mobiles robustes permettent, par rapport aux moyennes mobiles
classiques, de réduire les révisions et de mieux modéliser les points de
retournement autour de chocs (par exemple pendant la crise financière de
2008 ou le COVID-19), sans dégrader les estimations lorsqu'aucun choc
n'est observé. Les modélisations utilisées pour les méthodes
non-linéaires robustes ne permettent pas d'extraire une composante
tendance-cycle satisfaisante pour l'analyse conjoncturelle, avec parfois
des révisions importantes.

\section{Méthodologie}\label{sec-methodo}

L'hypothèse de base utilisée dans les méthodes de décomposition de
séries temporelles est que la série chronologique observée, \(y_t\),
peut être décomposée en une composante de signal \(\mu_t\) et une
composante erratique \(\varepsilon_t\) (appelée composante irrégulière)
: \[
f(y_t)=\mu_t+\varepsilon_t
\] où \(f\) désigne une transformation appropriée (généralement
logarithmique ou aucune transformation). Pour simplifier les notations
ultérieures, \(y_t\) désignera la série observée transformée. La
composante de bruit \(\varepsilon_t\) est généralement supposée être un
bruit blanc. En supposant que la série chronologique initiale est
désaisonnalisée (ou sans saisonnalité), le signal \(\mu_t\) représente
la tendance (variations sur une longue période) et le cycle (mouvements
cycliques superposés à la tendance à long terme), estimés ici
conjointement et appelé tendance-cycle \(TC_t\). L'estimation de la
composante tendance-cycle permet faciliter l'analyse des retournements
conjoncturels dans le cycle classique (également appelé cycle des
affaires)\footnote{ Dans cet article, on ne cherche donc pas à estimer
  séparément la tendance et le cycle, associée à des méthodes de
  filtrage de type Hodrick-Prescott ou Baxter-King, qui sont plutôt
  utilisées pour analyser les points de retournement dans le cycle de
  croissance (voir \textcite{ferrara2009FR} pour une description des
  différents cycles économiques).}. On parle de point de retournement
lorsque l'on passe d'une phase de récession (diminutions successives) à
une phase d'expansion de l'économie (augmentations successives) : on
parle alors de redressement ou, dans le cas contraire, de
ralentissement. Même si plusieurs formules peuvent être utilisées pour
définir ces phases, la définition de \textcite{Zellner1991} est
généralement utilisée pour déterminer les points de retournement : on a
un point de retournement à la date \(t\) lorsque
\(TC_{t-3}\geq TC_{t-2}\geq TC_{t-1}<TC_t\leq TC_{t+1}\) (redressement)
ou \(TC_{t-3}\leq TC_{t-2}\leq TC_{t-1}>TC_t\geq TC_{t+1}\)
(ralentissement).

La composante tendance-cycle est généralement estimée de manière locale.
En effet, autour d'un voisinage \(h\) de \(t\), cette composante peut
être approximée localement par un polynôme de degré \(d\)~: \[
TC_{t+i} = \sum_{j=0}^d\beta_j{i}^j+\xi_{t+i}\quad\forall i\in\{-h,-h+1,\dots,t+h\}
\] avec \(\xi_t\) un processus stochastique non corrélé avec
\(\varepsilon_t\) représentant l'erreur d'approximation. Même si
certains articles modélisent \(\xi_t\) et \(\varepsilon_t\) séparément
\autocite[voir par exemple][]{GrayThomson2002}, une hypothèse
habituelle, utilisée dans cet article, est de rassembler \(\xi_t\) et
\(\varepsilon_t\)\footnote{ Cela revient à supposer que le biais
  d'approximation de la tendance-cycle par un polynôme local est nul.}.
Ainsi, la tendance-cycle \(TC_t\) est considérée comme déterministe et
modélisée localement comme une tendance polynomiale de degré \(d\). Les
coefficients \((\beta_0,\dots,\beta_d)\) peuvent être estimés par la
méthode des moindres carrés pondérés. L'estimation \(\hat \beta_0\)
fournit l'estimation du cycle de tendance \(\widehat{TC}_t\) et on peut
montrer que cela équivaut à appliquer une moyenne mobile. Cette moyenne
mobile est généralement symétrique (\(\widehat{TC}_t\) est estimée en
utilisant autant d'observations avant et après \(t\)) mais pour
l'estimation des derniers points (lorsque l'on ne peut pas utiliser
autant d'observations avant et après \(t\)) il est nécessaire de
s'appuyer sur des moyennes mobiles \emph{ad hoc} qui sont asymétriques.

L'estimation de \(\widehat{TC}_t\) étant faite par un méthode linéaire,
elle est sensible à la présence de points atypiques qui peuvent
notamment entraîner des révisions importantes. C'est par exemple le cas
pendant la crise du COVID-19 qui a conduit Statistiques Canada et
Australian Bureau of Statistics à suspendre la publication des
tendances-cycles pendant cette période. Lors de périodes de fortes «
turbulences » (comme la crise du COVID-19) où les points atypiques sont
importants, les estimations directes de la tendance-cycle peuvent être
biaisées. Comme notamment discuté par \textcite{matthewssapw4},
différentes stratégies peuvent alors être adoptées :

\begin{itemize}
\item
  Ne pas publier les estimations pendant cette période : c'est ce qui
  est fait par l'Australian Bureau of Statistics pour la période du
  COVID-19 en ne publiant pas d'estimation tendance-cycle entre avril
  2020 et mars 2022.
\item
  Estimer les points atypiques avec un modèle RegARIMA, estimer la
  tendance-cycle sur la série corrigée et « réintroduire » les effets
  corrigés sur la tendance-cycle. L'inconvénient de cette approche est
  que l'estimation des points atypiques dépend du modèle ARIMA utilisé
  et de la période d'estimation. Si le modèle est réestimé à chaque
  nouvelle observation, la tendance-cycle peut donc potentiellement être
  révisée pendant de nombreuses périodes. Par exemple,
  \textcite{JMS2018DL} montrent que dans la majorité des cas il faut
  attendre au moins 3 ans pour que l'estimation du coefficient associé à
  une rupture converge. C'est l'approche privilégiée par l'Australian
  Bureau of Statistics lors de ruptures dans la tendance
  \autocite{abs2003}.
\item
  Estimer la tendance-cycle sur l'ensemble de la période et remplacer
  les données estimées après la rupture par celles estimées en
  commençant la série après la rupture (série segmentée d'un côté).
  L'inconvénient de cette approche est que les estimations avant la
  rupture vont être biaisées par la présence du choc et que celles après
  la rupture reposent sur des moyennes mobiles asymétriques (avec un
  biais plus important que les moyennes mobiles symétriques utilisées
  pour les estimations finales, et peuvent créer un déphasage,
  c'est-à-dire un décalage dans les points de retournement).
\item
  Diviser la série en deux au niveau de la rupture et estimer la
  tendance-cycle sur chaque segment (série segmentée des deux côtés).
  L'inconvénient de cette approche est que les estimations avant la
  rupture et après la rupture reposent sur des moyennes mobiles
  asymétriques (avec les mêmes inconvénients que précédemment). C'est
  l'approche privilégiée par Statistique Canada pendant le COVID-19
  \autocite{matthewssapw4} et celle qui était par l'Australian Bureau of
  Statistics lors des premières publications des estimations de la
  tendance-cycle \autocite{abs_smoothing_1987}.
\end{itemize}

L'objectif de cet article est d'étudier des approches alternatives à
celles de \textcite{matthewssapw4}. D'une part en étudiant les
estimations en temps réel de méthodes non-linéaires robustes à la
présence de points (section~\ref{sec-methodes-robustes}) ; d'autre part
en les comparant aux estimations des moyennes mobiles classiques
(section~\ref{sec-methodes-lineaires}) et en proposant une méthode de
construction de moyennes mobiles linéaires robustes
(section~\ref{sec-constr-mm}). Nous montrons également comment
construire des intervalles de confiance pour les estimations issues de
moyennes mobiles (section~\ref{sec-ic}), ce qui permet de valider le
choix de construction d'une moyenne mobile linéaire robuste.

\subsection{Méthodes robustes}\label{sec-methodes-robustes}

Dans cette étude nous étudions six méthodes robustes d'estimation locale
de la moyenne implémentées dans la fonction
\texttt{robfilter::robreg.filter()} \autocite{robfilter}. Dans le
package \texttt{robfilter}, la tendance-cycle est supposée être
localement linéaire : \[
y_{t+i}=\underbrace{\beta_{0,t}+\beta_{1,t}i}_{TC_{t+i}}+\varepsilon_{t,i}
\] On a donc \(\widehat{TC}_t = \hat\beta_{0,t}\) et l'on note
\(r_{t+i}= y_{t+i}- (\beta_{0,t}+ \beta_{1,t}i)\).

Les différentes méthodes étudiées sont :

\begin{itemize}
\tightlist
\item
  Médiane mobile (MED) :
\end{itemize}

\[
\widehat{TC}_t =
\underset{i=-h,\dots,h}{\med}y_{t+i}.
\]

\begin{itemize}
\item
  Régression médiane répétée --- \emph{Repeated Median} (RM)
  \autocite{rm} \[
  \hat{\beta}_{1,t}=
  \underset{i=-h,\dots,h}{\med}
  \left\{ \underset{i\ne j}{\med}\frac{y_{t+i}-y_{t+j}}{i-j} \right\}
  \] et \[
  \widehat{TC}_t=
  \underset{i=-h,\dots,h}{\med}
  \left\{ y_{t+i}-i\hat{\beta}_{1,t} \right\}.
  \]
\item
  Régression des moindres carrés médians --- \emph{Least Median of
  Squares} (LMS) \autocite{lms} : \[
  (\widehat{TC}_t,\hat\beta_{1,t})=
  \underset{\beta_{0,t},\beta_{1,t}}{\argmin}
  \left\{ \underset{i=-h,\dots,h}{\med}r^2_{t+i} \right\}.
  \]
\item
  Régression des moindres carrés élagués --- \emph{Least Trimmed
  Squares} (LTS) \autocite{lts} : \[
  (\widehat{TC}_t,\hat\beta_{1,t})=
  \underset{\beta_{0,t},\beta_{1,t}}{\argmin}
  \left\{ \sum_{i=-h}^h r^2_{t+i} \right\}.
  \]
\item
  Régression des moindres quartiles différenciés --- \emph{Least
  Quartile Difference} (LQD) \autocite{lqd} : \[
  \hat\beta_{1,t}=\underset{\beta_{1,t}}{\argmin}
  Q_{2h+1}(r_{t-h},\dots,r_{t+h})
  \] avec \[
  Q_{2h+1}(r_{t-h},\dots,r_{t+h}) = \left\{
  |r_i-r_j|;i<j
  \right\}_{\binom{h_p}{2}:\binom{2h+1}{2}}
  \] le quantile d'ordre \(\binom{h_p}{2}\) sur les \(\binom{2h+1}{2}\)
  éléments de l'ensemble \(\{|r_i-r_j|;i<j\}\) et \(h_p=[(2h+1+p+1)/2]\)
  avec \(p=1\) le nombre de régresseurs (hors constante). La fonction
  objectif ne dépendant pas de la constante, elle est estimée a
  posteriori par exemple en utilisant la formule : \[
  \widehat{TC}_t =
  \underset{i=-h,\dots,h}{\med}y_{t+i}-\hat \beta_{1,t}i.
  \]
\item
  Régression profonde --- \emph{Deepest Regression} (DR)
  \autocite{DeepestRegression} \[
  (\widehat{TC}_t,\hat\beta_{1,t})=\underset{\tilde\beta_{0},\tilde\beta_{1}}{\argmax}\left\{ rdepth(\tilde\beta_{0},\tilde\beta_{1}) \right\}
  \] où la profondeur de la régression --- \emph{regression depth}
  (\(rdepth\)) --- d'un ajustement \((\tilde\beta_{0},\tilde\beta_{1})\)
  est définie comme : \[
  rdepth(\tilde\beta_{0},\tilde\beta_{1}) = \underset{-h\leq i\leq h}{\min}
  \left\{ \min \left\{ L^+(i) + R^-(i);R^+(i) + L^-(i) \right\}\right\}
  \] avec : \[
  \begin{cases}
  L^+(i) = L^+_{\tilde\beta_{0},\tilde\beta_{1}}(i) = cardinal \left\{ j\in\{-h,\dots,i\} : r_j(\tilde\beta_{0},\tilde\beta_{1})\leq 0 \right\} \\
  R^-(i) = R^-_{\tilde\beta_{0},\tilde\beta_{1}}(i) = cardinal \left\{ j\in\{i+1,\dots,h\} : r_j(\tilde\beta_{0},\tilde\beta_{1})< 0 \right\}
  \end{cases}
  \] et \(L^-(i)\) et \(R^+(i)\) définis de manière analogue.
\end{itemize}

Pour les estimations intermédiaires (i.e.~: lorsque pour estimer la
tendance-cycle à la date \(t\) on ne possède pas \(h\) observations
avant ou après \(t\)), le package \texttt{robfilter} extrapole les
données à partir du dernier modèle estimé. C'est-à-dire que si les
données observées sont \(y_1, \dots, y_n\), alors pour
\(q\in \{1,\dots, 6\}\) : \[
\widehat{TC}_{n-h+q} = \widehat{TC}_{n-h}+q\times\hat\beta_{1,n-h}.
\] Pour la médiane mobile, cela revient à prolonger par la dernière
valeur connue. Cette méthode n'est pas pertinente pour l'analyse
conjoncturelle puisque les points de retournement sont détectés avec
\(h\) périodes de retard. Une approche alternative consiste à appliquer
les mêmes modèles mais en n'utilisant que les observations disponibles
(pour estimer \(\widehat{TC}_{n-h+1}\) on utilise les \(2h\)
observations entre \(n-2h+1\) et \(n\), \ldots, et pour estimer
\(\widehat{TC}_{n}\) on utilise les \(h+1\) observations entre \(n-h\)
et \(n\)). Cette approche n'est pas implémentée dans \texttt{robfilter}
mais peut être obtenue en ajoutant \(h\) valeurs manquantes au début et
à la fin de la série : c'est ce qui sera effectué dans cet article.

Dans le cadre de l'analyse conjoncturelle, il n'est généralement pas
plausible de modéliser, autour des points de retournement, une tendance
de degré 1\footnote{ Dans l'utilisation de méthodes robustes, il est
  courant de localement modéliser une tendance de degré 1. Par exemple,
  dans la méthode de lissage Loess, \textcite{loess} recommande
  d'utiliser une tendance de degré 1 qui permet d'avoir un «~bon
  compromis entre facilité de calcul et besoin de flexibilité ~». Idem,
  dans la méthode de désaisonnalisation STL (\emph{Seasonal-Trend
  decomposition based on Loess}), \textcite{cleveland90} indiquent :
  «~Il est raisonnable de prendre \(d=1\) si le modèle sous-jacent des
  données présente une légère courbure. Mais si la courbure est
  importante, par exemple avec des pics et des creux, il est préférable
  d'opter pour \(d=2\).~»}. Il n'est toutefois pas possible de changer
cela dans le package \texttt{robfilter} et il n'existe pas d'autre
package permettant d'appliquer simplement toutes ces méthodes. Même si
cette modélisation n'est pas toujours optimale, elle reste plausible
dans la majorité des cas et la facilité d'utilisation de
\texttt{robfilter} font de ces méthodes un bon point de comparaison aux
moyennes mobiles linéaires classiquement utilisées pour l'estimation de
la tendance-cycle et présentées dans la section suivante. Toutefois, les
méthodes LTS et LMS sont implémentées dans le package \texttt{MASS}
\autocite{MASS}\footnote{ Nous n'avons pas trouvé d'autres fonctions
  permettant d'utiliser les autres méthodes, même si elles pourraient
  être réimplémentées facilement.}, il est possible de réimplémenter le
processus d'estimation locale avec une tendance de degré 2, ce qui a été
fait dans l'annexe \ref{sec-an-lms-lts}.

\subsection{Moyennes mobiles linéaires
classiques}\label{sec-methodes-lineaires}

\subsubsection{Moyenne mobile de Henderson et de Musgrave}\label{sec-lp}

Les moyennes mobiles classiques peuvent être obtenues par analogie avec
la régression polynomiale locale. En reprenant les notations de
\textcite{proietti2008}, on suppose que notre série chronologique
\(y_t\) peut être décomposée en : \[
y_t=TC_t+\varepsilon_t,
\] où \(TC_t\) est la tendance-cycle et
\(\varepsilon_{t}\overset{i.i.d}{\sim}\mathcal{N}(0,\sigma^{2})\) est le
bruit. La tendance-cycle \(TC_t\) est localement approchée par un
polynôme de degré \(d\), de sorte que dans un voisinage \(h\) de \(t\)
on a \(TC_t\simeq m_{t}\) avec : \[
\forall j\in\left\{-h,-h+1,\dots,h\right\},\:
y_{t+j}=m_{t+j}+\varepsilon_{t+j},\quad m_{t+j}=\sum_{i=0}^{d}\beta_{i,t}j^{i}.
\] En notation matricielle :
\begin{equation}\phantomsection\label{eq-lpp}{
\underbrace{\begin{pmatrix}y_{t-h}\\
y_{t-(h-1)}\\
\vdots\\
y_{t}\\
\vdots\\
y_{t+(h-1)}\\
y_{t+h}
\end{pmatrix}}_{\boldsymbol y_t}=\underbrace{\begin{pmatrix}1 & -h & h^{2} & \cdots & (-h)^{d}\\
1 & -(h-1) & (h-1)^{2} & \cdots & (-(h-1))^{d}\\
\vdots & \vdots & \vdots & \cdots & \vdots\\
1 & 0 & 0 & \cdots & 0\\
\vdots & \vdots & \vdots & \cdots & \vdots\\
1 & h-1 & (h-1)^{2} & \cdots & (h-1)^{d}\\
1 & h & h^{2} & \cdots & h^{d}
\end{pmatrix}}_{\boldsymbol X}\underbrace{\begin{pmatrix}\beta_{0,t}\\
\beta_{1,t}\\
\vdots\\
\vdots\\
\vdots\\
\vdots\\
\beta_{d,t}
\end{pmatrix}}_{\boldsymbol \beta_t}+\underbrace{\begin{pmatrix}\varepsilon_{t-h}\\
\varepsilon_{t-(h-1)}\\
\vdots\\
\varepsilon_{t}\\
\vdots\\
\varepsilon_{t+(h-1)}\\
\varepsilon_{t+h}
\end{pmatrix}}_{\boldsymbol \varepsilon_t}.
}\end{equation}

L'estimation des paramètres \(\boldsymbol \beta_t\) peut être obtenue
moindres carrés pondérés --- \emph{weighted least squares} (WLS) --- à
partir d'un ensemble de poids \((\kappa_j)_{-h\leq j \leq h}\) appelés
noyaux. En notant \(\boldsymbol K=diag(\kappa_{-h},\dots,\kappa_{h})\)
il vient
\(\hat{\boldsymbol\beta_t}=(\transp{\boldsymbol X}\boldsymbol K\boldsymbol X)^{-1}\transp{\boldsymbol X}\boldsymbol K\boldsymbol y_t.\)
Avec
\(\boldsymbol e_1=\transp{\begin{pmatrix}1 &0 &\cdots&0 \end{pmatrix}}\),
l'estimation de la tendance-cycle est :
\begin{equation}\phantomsection\label{eq-mmsym}{
\widehat{TC}_t=\widehat{\beta}_{0,t}=\transp{\boldsymbol e_{1}}\hat{\boldsymbol \beta_t}=\transp{\boldsymbol \theta}\boldsymbol y_t=\sum_{j=-h}^{h}\theta_{j}y_{t-j}\text{ avec }\boldsymbol \theta=\boldsymbol K\boldsymbol X(\transp{\boldsymbol X}\boldsymbol K\boldsymbol X)^{-1}\boldsymbol e_{1}.
}\end{equation} En somme, l'estimation de la tendance \(\hat{m}_{t}\)
est obtenue en appliquant une moyenne mobile symétrique
\(\boldsymbol \theta\) à \(y_t\).

On retrouve la moyenne mobile de Henderson avec \(d=2\) (ou \(d=3\)) et
en utilisant les noyaux~: \[
\kappa_{j}=\left[1-\frac{j^2}{(h+1)^2}\right]
\left[1-\frac{j^2}{(h+2)^2}\right]
\left[1-\frac{j^2}{(h+3)^2}\right].
\]

Pour le cas asymétrique, \textcite{proietti2008} proposent une méthode
générale pour construire les filtres asymétriques qui permet de faire un
compromis biais-variance. Il s'agit d'une généralisation des filtres
asymétriques de \textcite{musgrave1964set} (utilisés dans l'algorithme
de désaisonnalisation X-13ARIMA-SEATS). En récrivant
l'équation~\ref{eq-lpp} :
\begin{equation}\phantomsection\label{eq-lpgeneralmodel}{
\boldsymbol y_t=\begin{pmatrix}\boldsymbol U &\boldsymbol Z\end{pmatrix}
\begin{pmatrix}\boldsymbol \gamma_t \\ \boldsymbol \delta_t\end{pmatrix}+
\boldsymbol \varepsilon_t = 
\boldsymbol U\boldsymbol \gamma_t + \boldsymbol Z \boldsymbol \delta_t +\boldsymbol \varepsilon_t,
\quad
\boldsymbol \varepsilon_t\sim\mathcal{N}(0,\boldsymbol D).
}\end{equation} où \([\boldsymbol U,\boldsymbol Z]\) est de rang plein
et forme un sous-ensemble des colonnes de \(\boldsymbol X\). L'objectif
est de trouver une moyenne mobile asymétrique
\(\boldsymbol \theta^{(a)}\) qui minimise l'erreur quadratique moyenne
de révision (à la moyenne mobile symétrique \(\boldsymbol \theta\)) sous
certaines contraintes. Ces contraintes sont représentées par la matrice
\(\boldsymbol U=\transp{\begin{pmatrix}\transp{\boldsymbol U_{p}}&\transp{\boldsymbol U_{f}}\end{pmatrix}}\)
:
\(\transp{\boldsymbol U_p}\boldsymbol \theta^{(a)}=\transp{\boldsymbol U}\boldsymbol \theta\)
(avec \(\boldsymbol U_p\) la matrice \((h+q+1)\times (d+1)\) qui
contient les observations de la matrice \(\boldsymbol U\) connues lors
de l'estimation par le filtre asymétrique). C'est ce qui est implémenté
dans la fonction \texttt{rjd3filters::mmsre\_filter()} du package R
\texttt{rjd3filters} \autocite{rjd3filters}.

Lorsque \(\boldsymbol U\) correspond aux \(d^*+1\) premières colonnes de
\(\boldsymbol X\), \(d^*<d\), la contrainte consiste à reproduire des
tendances polynomiales de degré \(d^*\). Cela introduit du biais (sur
les tendances de degré \(d^*+1\) à \(d\)) mais réduit la variance. Le
filtre de Musgrave se retrouve en modélisant une tendance localement
linéaire (\(d=1\)), en imposant que \(\theta^{(a)}\) préserve les
constantes (\(d^*=0\)) et en prenant le filtre d'Henderson comme filtre
symétrique. C'est-à-dire que l'on a
\(\boldsymbol U=\transp{\begin{pmatrix}1&\cdots&1\end{pmatrix}}\),
\(\boldsymbol Z=\transp{\begin{pmatrix}-h&\cdots&+h\end{pmatrix}}\),
\(\boldsymbol \delta_t=\delta_{1,t}\),
\(\boldsymbol D=\sigma^2\boldsymbol I\). Ce filtre dépend du rapport
\(\lvert\delta_{1,t}/\sigma\rvert\) (modélisant le biais), qui est à
fixer par l'utilisateur. En supposant que la tendance est linéaire et le
biais constant (\(\delta_{1,t}=\delta_1\)), ce rapport est lié à l'I-C
ratio
\(R=\frac{\bar{I}}{\bar{C}}=\frac{\sum\lvert I_t-I_{t-1}\rvert}{\sum\lvert C_t-C_{t-1}\rvert}\)
(et l'on a \(\delta_1/\sigma=2/(R\sqrt{\pi})\)), qui est notamment
utilisé dans X-11 pour déterminer la longueur de la moyenne mobile de
Henderson à utiliser. Pour des données mensuelles :

\begin{itemize}
\item
  Si le rapport est élevé (\(3,5< R\)), un filtre symétrique à 23 termes
  est utilisé (pour éliminer plus de bruit) et le rapport \(R=4,5\) est
  utilisé dans X-11 pour définir le filtre de Musgrave.
\item
  Si le rapport est faible (\(R<1\)), un filtre symétrique à 9 termes
  est utilisé et le rapport \(R=1\) est utilisé dans X-11 pour définir
  le filtre de Musgrave.
\item
  Dans le cas contraire (\(1\leq R \leq 3,5\), la plupart des cas), un
  filtre symétrique à 13 termes est utilisé et le rapport \(R=3,5\) est
  utilisé dans X-11 pour définir le filtre de Musgrave.
\end{itemize}

Dans cet article, seules des séries mensuelles sont étudiés et le filtre
de Henderson à 13 termes sera utilisé comme filtre symétrique. Par
simplification et par cohérence avec ce qui est utilisé dans X-11, le
ratio \(\delta_1/\sigma\) sera fixé de sorte à avoir \(R=3,5\). C'est
également ce qui est fait par l'Australian Bureau of Statistics
\autocite{abs2003}. Toutefois, comme notamment montré dans
\textcite{jos2024AQLT}, une paramétrisation locale de ce paramètre
pourrait être préférée.

Par simplification, dans cette étude on parle d'estimation via la
moyenne mobile de Henderson pour désigner l'utilisation des moyennes
mobiles de Henderson pour l'esitmation finale et des moyennes mobiles de
Musgrave pour les estimations intermédiaires. Les coefficients des
différentes moyennes mobiles étudiées dans cet article sont représentés
dans l'annexe \ref{sec-annexes-mm}.

\subsubsection{Filtres cascade}\label{filtres-cascade}

Les moyennes mobiles étant des opérateurs linéaires, ils sont sensibles
à la présence de points atypiques. Du fait de leur présence,
l'application directe des méthodes peut donc conduire à des estimations
biaisées alors que les méthodes de désaisonnalisation (comme la méthode
X-13ARIMA-SEATS) ont un module de correction des points atypiques. Par
ailleurs, comme notamment montré par \textcite{dagum1996new}, le filtre
symétrique final utilisé par X-11 pour extraire la tendance-cycle (et
donc celui indirectement utilisé lorsqu'on applique les méthodes sur les
séries désaisonnalisées) laisse passer environ 72 \% des cycles de 9 ou
10 mois (généralement associés à du bruit plutôt qu'à la
tendance-cycle). Les filtres asymétriques finaux amplifient même les
cycles de 9 ou 10 mois. Cela peut avoir pour conséquence l'introduction
d'ondulations indésirables, c'est-à-dire la détection de faux points de
retournement. Ce problème est réduit par la correction des points
atypiques (ces cycles étant considérés comme de l'irrégulier). C'est
ainsi que le \emph{Nonlinear Dagum Filter} (NLDF) a été développé et
consiste à :

\begin{enumerate}
\def\labelenumi{\arabic{enumi}.}
\item
  appliquer l'algorithme de correction des points atypiques de X-11 sur
  la série désaisonnalisée, puis la prolonger par un modèle ARIMA ;
\item
  effectuer une nouvelle correction des points atypiques en utilisant un
  seuil bien plus strict et appliquer ensuite le filtre symétrique de 13
  termes. En supposant une distribution normale cela revient à modifier
  48 \% des valeurs de l'irrégulier.
\end{enumerate}

Les \emph{cascade linear filter} \autocite[CLF,][]{clf}, correspondent à
une approximation des NLDF en utilisant un filtre de 13 termes et
lorsque les prévisions sont obtenus à partir d'un modèle ARIMA(0,1,1) où
\(\theta=0,40.\) De manière similaire, des moyennes mobiles
asymétriques, \emph{Asymmetric Linear Filter} (ALF), peuvent être
construites.

C'est la moyenne mobile CLF qui est utilisée par Statistique Canada pour
l'estimation de la tendance-cycle\footnote{ Voir par exemple
  \url{https://www.statcan.gc.ca/fr/quo/bdd/tendance-cycle}.}. Pour
l'estimation en temps réel, Statistique Canada utilise la méthode «
couper-et-normaliser » qui consiste à recalculer les poids, à partir de
la moyenne mobile symétrique, en n'utilisant que les poids associées aux
observations disponibles (couper) et en renormalisant pour que la somme
des coefficients soit égale à 1 (normaliser). C'est l'approche qui sera
utilisé dans cet article (plutôt que l'utilisation des ALF).

\begin{remark}
Les CLF ne préservent, localement, que les droites (tendances de degré
1) et les moyennes mobiles asymétriques associées que les constantes
(tendances de degré 0). En revanche la moyenne mobile de Henderson
préserve les tendances locales de degré 3 et les moyennes mobiles de
Musgrave que les constantes.
\end{remark}

\subsection{Construction de moyennes mobiles
robustes}\label{sec-constr-mm}

Une autre façon de construire des moyennes mobiles robustes aux points
atypiques est d'exploiter l'approche polynomiale locale présentée dans
la section~\ref{sec-lp}. Pour la construire la moyenne mobile utilisée
pour l'estimation finale de la tendance-cycle, il suffit pour cela
d'ajouter dans l'équation~\ref{eq-lpp} un ou plusieurs régresseurs
\(\boldsymbol O_t\) pour prendre en compte l'effet du point atypique
modélisé : \[
y_{t+j}=\sum_{i=0}^{d}\beta_{i,t}j^{i} + \boldsymbol O_{t+j} \boldsymbol\zeta_t+\varepsilon_{t+j}.
\]

On a alors : \[
\widehat{TC}_t=\transp{\boldsymbol \theta_t^{(r)}}\boldsymbol y_t=\sum_{j=-h}^{h}\theta_{j,t}^{(r)}y_{t-j}
\] avec \[
\boldsymbol \theta_t^{(r)}=\boldsymbol K\begin{pmatrix}\boldsymbol X &  \mathcal O_{t}\end{pmatrix}\left(\transp{\begin{pmatrix}\boldsymbol X &  \mathcal O_{t}\end{pmatrix}}\boldsymbol K \begin{pmatrix}\boldsymbol X &  \mathcal O_{t}\end{pmatrix}\right)^{-1}\boldsymbol e_{1}
\text{ et }
\mathcal O_{t} = \begin{pmatrix}
\boldsymbol O_{t-h} \\ \vdots\\\boldsymbol O_{t+h}
\end{pmatrix}.
\]

Le filtre de Henderson peut être obtenu par approximation locale d'un
polynôme de degré 2. Puisqu'il est symétrique, il préserve également les
tendances de degré 3 et peut également être obtenu par approximation
locale d'un polynôme de degré 3. En revanche, même si la moyenne mobile
\(\boldsymbol \theta_t^{(r)}\) est centrée (autant de points avant et
après \(t\) sont utilisés pour estimer \(\widehat{TC}_t\)), elle n'est
plus forcément symétrique (pour au moins un \(i\in\{1,\dots,h\}\) on a
\(\theta_{-i,t}^{(r)}\ne\theta_{i,t}^{(r)}\)). Pour construire une
moyenne mobile robuste associée au filtre de Henderson, le choix entre
modélisation d'un polynôme local de degré 2 (\(d=2\)) ou 3 (\(d=3\))
aura donc un impact sur la moyenne mobile finale. Dans cet article nous
utiliserons par convention la valeur \(d=3\) afin de préserver sans
biais les mêmes tendances que la moyenne mobile de Henderson.

Pour la construction des moyennes mobiles asymétriques, le même principe
est appliqué à l'équation~\ref{eq-lpgeneralmodel} en estimant sans biais
\(\boldsymbol\zeta_t\) : \[
\boldsymbol y_t=\begin{pmatrix}\boldsymbol U & \mathcal O_{t}\end{pmatrix}\begin{pmatrix}\boldsymbol \gamma_t \\ \boldsymbol\zeta_t \end{pmatrix}+\boldsymbol Z\boldsymbol \delta_t+\boldsymbol \varepsilon_t,\quad
\boldsymbol \varepsilon_t\sim\mathcal{N}(\boldsymbol 0,\boldsymbol D).
\] La moyenne mobile asymétrique \(\boldsymbol \theta^{(a)(r)}_t\) est
trouvée par minimisation de l'erreur quadratique moyenne de révision (à
la moyenne mobile \(\boldsymbol \theta_t^{(r)}\)) sous contraintes. Ces
contraintes sont représentées par la matrice
\(\begin{pmatrix}\boldsymbol U & \mathcal O_{t}\end{pmatrix} = \begin{pmatrix}\boldsymbol U_{p} & \mathcal O_{p,t} \\\boldsymbol U_{f} & \mathcal O_{f,t} \end{pmatrix}\)
: \[
\transp{\begin{pmatrix}\boldsymbol U_{p} & \mathcal O_{p,t} \end{pmatrix}}\boldsymbol \theta^{(a)}=\transp{\begin{pmatrix}\boldsymbol U & \mathcal O_{t}\end{pmatrix}} \boldsymbol \theta_t^{(r)}.
\]

Par convention, lorsque \(\mathcal O_{t}\) ou \(\mathcal O_{p,t}\) est
la matrice nulle, on garde la modélisation de la section~\ref{sec-lp}
(sans introduction de régresseur supplémentaire).

Dans cette étude, les deux types de points atypiques les plus couramment
rencontrés dans l'analyse de données macroéconomiques sont étudiés :

\begin{itemize}
\item
  Les points atypiques additifs (AO, \emph{additive outlier}) : un choc
  ponctuelle à une date particulière puis un retour à la normal (grève,
  mesure exceptionnelle, erreur de mesure par exemple liée à de la non
  réponse, etc.). Le choc affecte donc l'irrégulier et ne devrait pas
  avoir d'impact sur la tendance-cycle. Si le choc apparaît à la date
  \(t_0\) il peut être modélisé par le régresseur
  \(O_t^{AO}=\1_{t=t_0}\) qui vaut \(1\) à la date \(t_0\) et \(0\)
  sinon.\\
  Si l'on souhaite que le choc soit affecté à la tendance-cycle et non
  pas à l'irrégulier (par exemple pendant la période du COVID-19), il
  faut utiliser le régresseur \(O_{t+j}=O_{t+j}^{AO}=\1_{t+j= t_0}\)
  lorsque \(t< t_0\) ou \(t=t_0+h\) (estimations des points avant le
  choc et dernière moyenne mobile modifiée) et
  \(O_{t+j}=1-O_{t+j}^{AO}=\1_{t+j\ne t_0}\) lorsque
  \(t_0\leq t < t_0+h\) (estimations des points après le choc sauf
  dernière moyenne mobile modifiée). Une distinction est faite pour
  \(t=t_0+h\) car dans ce cas le régresseur \(\1_{t+j\ne t_0}\) est égal
  à \(\1_{t+j> t_0}\) ce qui revient à créer une rupture en niveau (cf
  \emph{infra}) et introduit un pic indésirable à l'estimation en
  \(t_0+h\).
\item
  Les ruptures en niveau (LS, \emph{level shift}) : un changement
  soudain et durable du niveau moyen de la série (choc structurel,
  changement de politique économique, etc.). Le choc affecte donc la
  tendance-cycle et ne devrait pas avoir d'impact sur l'irrégulier. Si
  le choc est à la date \(t_0\) il peut être modélisé par le régresseur
  \(O_{t+j}=1-\1_{t+j< t_0}\) si \(t\leq t_0\) (l'estimation de
  \(\beta_{0,t}\) ne prend pas en compte le choc en niveau) et
  \(O_{t+j}=\1_{t+j\geq t_0}\) si \(t_0<t\) (l'estimation de
  \(\beta_{0,t}\) prend en compte le choc en niveau).
\end{itemize}

Cette méthode a été implémentée dans la fonction
\texttt{publishTC::henderson\_robust\_smoothing()} du package
\texttt{publishTC} \autocite{publishTC} qui permet de modéliser des
chocs ponctuels (affectés à l'irrégulier ou à la tendance-cycle) et des
ruptures en niveau (affectées à la tendance-cycle).

L'inconvénient de cette approche, par rapport aux méthodes robustes
présentées dans la section~\ref{sec-methodes-robustes}, est que cela
suppose de connaître la date et la nature du point atypique. Une
approche en deux temps peut être utilisée :

\begin{enumerate}
\def\labelenumi{\arabic{enumi}.}
\item
  Partir sur un a priori basé sur une information économique (ex :
  baisse attendue de l'économie au moment des confinements durant le
  COVID-19, hausse attendue suite à une mise en place d'une
  politique\ldots) ou sur des modèles statistiques (comme le module de
  détection des points atypiques de X-13ARIMA-SEATS basé sur un modèle
  RegARIMA).
\item
  Valider la modélisation retenue en comparant l'intervalle de confiance
  de l'estimation de la tendance-cycle avec les moyennes mobiles de
  Henderson et de Musgrave avec l'estimation en utilisant les moyennes
  mobiles robustes. La section~\ref{sec-ic} décrit la méthodologie pour
  la construction d'intervalles de confiance pour des estimations basées
  sur des moyennes mobiles.
\end{enumerate}

Toutefois, pour l'estimation en temps réel, sauf à avoir une information
extérieure, il est difficile de distinguer un \emph{level shift} (LS)
d'un ou plusieurs \emph{additive outlier} (AO) consécutifs. Le choix du
type de point atypique aura une influence forte sur l'estimation
(puisque les LS affectent la tendance-cycle et les AO l'irrégulier) et
passer d'une spécification à l'autre entraînera des révisions
importantes. Ces révisions seront même plus importantes que si, avant de
décider de la bonne modélisation à adopter, les premières estimations
avaient été faites avec des moyennes moyennes mobiles classiques.

Par rapport au préajustement des points atypiques par un modèle \emph{ad
hoc}, comme le module de pré-ajustement de X-13ARIMA-SEATS basé sur un
modèle RegARIMA, l'approche ici présentée a l'avantage de ne pas être
dépendante de l'identification d'un modèle et de la période d'estimation
(pas de biais lié à l'identification des paramètres ou à l'utilisation
d'une période d'estimation trop courte ou trop longue).

Dans la suite ces moyennes mobiles seront appelées moyennes mobiles de
Henderson ou de Musgrave robustes. Par simplification, nous appellerons
également moyenne mobile de Henderson robuste l'utilisation des moyennes
de mobiles de Henderson (pour l'estimation finale) et de Musgrave (pour
les estimations intermédiaires) robustes construits selon la même
méthodologie.

\subsection{Construction d'intervalles de confiance pour des moyennes
mobiles}\label{sec-ic}

Soit \(y_1,\dots,y_n\) une série chronologique observée. On suppose
qu'elle peut être décomposée en \[
y_t=\mu_t+\varepsilon_t,
\] où \(\mu_t\) est une composante déterministe à estimer et
\(\varepsilon_{t}\overset{i.i.d}{\sim}\mathcal{N}(0,\sigma^{2})\) est le
bruit.

Soit
\(\boldsymbol\theta = \begin{pmatrix}\theta_{-p},\dots,\theta_f\end{pmatrix}\)
une moyenne mobile permettant d'estimer la composante inobservable
\(\mu_t\) (dans cette étude la tendance-cycle \(TC_t\)) à partir de
\(y_t\). Cette estimation est donnée par
\(\hat \mu_t = \sum_{i=-p}^{+f}\theta_iy_{t+i}.\)

Un intervalle de confiance de \(\mathbb E[\hat{\mu}_t]\) au seuil
\(\alpha\) peut être calculé à partir de la formule :
\begin{equation}\phantomsection\label{eq-ic}{
I_t=\left[\hat{\mu}_t - |q_{\alpha/2}|\sqrt{\hat \sigma^2}\sqrt{\sum_{i=-p}^{+f}\theta_i^2};
 \hat{\mu}_t  + |q_{1-\alpha/2}|\sqrt{\hat \sigma^2}\sqrt{\sum_{i=-p}^{+f}\theta_i^2}\right]
}\end{equation} où \[
\hat\sigma^2=\frac{1}{(n-p-f)\left(1-2\theta_0^2+\sum_{i=-p}^{+f} \theta_i^2\right)}\sum_{t=p+1}^{n-f}(y_t-\widehat{\mu}_t)^2
\] et \(q_{\alpha/2}\) est le quantile d'ordre \(\alpha/2\) d'une
certaine loi de Student. C'est un intervalle de confiance de \(\mu_t\)
lorsque l'on a un estimateur sans biais de \(\mu_t\)
(\(\mathbb E[\hat{\mu}_t]= \sum_{i=-p}^{+f}\theta_iy_{t+i}= \mu_t\)), ce
qui n'est généralement pas le cas, mais le biais est négligeable lorsque
la fenêtre \(p+f+1\) est petite.

Ces formules se retrouvent par analogie avec la régression polynomiale
locale. En reprenant les notations de \textcite{Loader1999} pour la
régression polynomiale locale et en adaptant à l'utilisation de moyennes
mobiles, la variance \(\hat\sigma^2\) peut être estimée par la somme des
carrés des résidus normalisés : \[
\hat\sigma^2=\frac{1}{(n-p-f)-2\nu_1+\nu_2}\sum_{t=p+1}^{n-f}(y_t-\widehat{\mu}_t)^2.
\] \(n-p-f\) termes sont utilisés car avec la moyenne mobile
\(\boldsymbol \theta\) seulement \(n-p-f\) observations peuvent être
utilisées pour estimer \(\sigma^2\). \(\nu_1\) et \(\nu_2\) sont deux
définitions de degrés de liberté d'une estimation locale (généralisation
du nombre de paramètres d'un modèle paramétrique). Notons
\(\boldsymbol H\) la \emph{matrice chapeau} de taille \(n\times n\)
permettant de faire correspondre les données aux valeurs estimées : \[
\begin{pmatrix}
\hat{\mu}_1\\ \vdots \\ \hat{\mu}_n
\end{pmatrix} = \boldsymbol H \boldsymbol Y
\text{ avec }
\boldsymbol Y = \begin{pmatrix}
y_1\\ \vdots \\ y_n
\end{pmatrix}.
\] En considérant par convention que
\(\hat\mu_1=\dots=\hat{\mu}_p=\hat\mu_{n-f+1}=\dots=\hat\mu_n=0\)
(puisque l'on ne peut pas estimer ces quantités avec la moyenne mobile
\(\boldsymbol \theta\)), on a donc : \[
\boldsymbol H=\begin{pmatrix}
&&\boldsymbol0_{p\times n} \\
\theta_{-p} & \cdots & \theta_f  & 0 & \cdots\\
0 & \theta_{-p} & \cdots & \theta_f  & 0 & \cdots
\\ & \ddots &&&\ddots\\
0 &\cdots&0& \theta_{-p} & \cdots & \theta_f \\
&&\boldsymbol0_{f\times n} 
\end{pmatrix},
\] où \(\boldsymbol 0_{p\times n}\) est la matrice de taille
\(p\times n\) ne contenant que de zéros. On a : \[
\begin{cases}
\nu_1 =\tr (\boldsymbol H) = (n-p-f)\theta_0\\
\nu_2 = \tr (\transp{\boldsymbol H}\boldsymbol H) = (n-p-f) \sum_{i=-p}^{+f} \theta_i^2
\end{cases}.
\] Si les bruits \(\varepsilon_t\) sont indépendants et de variance
\(\sigma^2\), alors : \[
\V{y_t-\hat{\mu}_t}=\sigma^2 - 
2\underbrace{\cov{y_t}{\hat{\mu}_t}}_{=\theta_0\sigma^2} + 
\underbrace{\V{\hat{\mu}_t}}_{=\sigma^2\sum_{i=-p}^{+f} \theta_i^2}
\] et l'on a donc : \[
\E{\hat\sigma^2}=\sigma^2 + \frac{1}{(n-p-f)-2\nu_1+\nu_2}\sum_{t=p+1}^{n-f}(\E{\hat \mu_t}-\mu_t)^2.
\] L'estimateur \(\hat\sigma^2\) est donc sans biais si \(\hat\mu_t\)
l'est aussi.

La somme des carrés des résidus peut s'écrire sous forme quadratique :
\[
\sum_{t=p+1}^{n-f}(y_t-\widehat{\mu}_t)^2 =
\transp{\boldsymbol Y}\boldsymbol \Delta\boldsymbol Y 
\] avec
\({\boldsymbol Y} = \transp{\begin{pmatrix} y_1 & \cdots & y_n\end{pmatrix}}\)
et : \[
\boldsymbol \Delta = \transp{(\boldsymbol I - \boldsymbol H)}(\boldsymbol I - \boldsymbol H\boldsymbol),\,
\boldsymbol I = \begin{pmatrix}
&\boldsymbol0_{p\times n} \\
\boldsymbol 0_{(n-p-f)\times p} &\boldsymbol I_{n-p-f} & \boldsymbol 0_{(n-p-f)\times f}  \\
&\boldsymbol0_{f\times n} 
\end{pmatrix}
\] et \(\boldsymbol I_{n-p-f}\) la matrice identité de taille \(n-p-f.\)
On a donc : \[
\hat\sigma^2=\frac{1}{\tr{\boldsymbol \Delta}}\transp{\boldsymbol Y}\boldsymbol \Delta\boldsymbol Y.
\] Si les bruits \(\varepsilon_t\) sont normalement distribués
indépendants et \(\hat\sigma^2\), la distribution est : \[
\transp{\boldsymbol Y}\boldsymbol \Delta\boldsymbol Y\overset{\mathcal L}{=}\sigma^2\sum_{j=1}^n\lambda_jZ_j,
\] où les \(\lambda_j\) sont les valeurs propres de
\(\boldsymbol\Delta\) et les \(Z_j\) sont des lois indépendantes du
chi-deux à 1 degré de liberté. Il vient : \[
\begin{cases}
\E{\hat\sigma^2}=\sigma^2\frac{1}{\tr{\boldsymbol \Delta}}\underbrace{\sum_{j=1}^n\lambda_j}_{=\tr{\boldsymbol \Delta}}=\sigma^2\\
\V{\hat\sigma^2}=\sigma^4\frac{1}{\tr{(\boldsymbol \Delta)}^2}\sum_{j=1}^n2\lambda_j^2=2\sigma^4\frac{\tr(\boldsymbol \Delta^2)}{\tr(\boldsymbol \Delta)^2}
\end{cases}.
\] En notant
\(\nu = \tr(\boldsymbol \Delta)^2 / \tr(\boldsymbol \Delta^2)\), on a
donc : \[
\begin{cases}
\E{\nu\frac{\hat\sigma^2}{\sigma^2}}=\nu\\
\V{\nu\frac{\hat\sigma^2}{\sigma^2}}=2\nu
\end{cases}.
\] Les deux premiers moments de \(\nu \hat\sigma^2/\sigma^2\) sont donc
identiques à ceux d'un loi du chi-deux avec \(\nu\) degrés de liberté.
La loi \(\hat\sigma^2\) étant difficile à calculer dans le cadre de la
régression locale, on donc approximer sa distribution par une loi du
chi-deux.

Puisque \(\V{\mu_t}=\sigma^2\sum_{i=-p}^{+f} \theta_i^2\), on retrouve
la formule de l'intervalle de confiance de l'équation~\ref{eq-ic}~: \[
\frac{\hat\mu_t-\E{\mu_t}}{\sqrt{\V{\mu_t}}}=\frac{\hat\mu_t-\E{\mu_t}}{\sqrt{\sigma^2\sum_{i=-p}^{+f} \theta_i^2}}
\text{ et }
\frac{\hat\mu_t-\E{\mu_t}}{\sqrt{\hat\sigma^2}\sqrt{\sum_{i=-p}^{+f} \theta_i^2}}\sim\mathcal{T}(\nu).
\] Le numérateur de \(\nu\) se calcule facilement puisque l'on a : \[
\tr(\boldsymbol \Delta) = (n-p-f)\left(1-2\theta_0+\sum_{i=-p}^{+f} \theta_i^2\right).
\] En revanche il est difficile d'avoir une formule simplifiée pour le
dénominateur \(\tr(\boldsymbol \Delta^2).\) Ce dernier peut être calculé
en reconstruisant la matrice \(\boldsymbol \Delta\) (produit de matrices
de taille \(n\times n\)) ou, comme proposé dans cet article, par le
produit d'une matrice de taille \(1\times (p+f+1)\) et d'une matrice de
taille \((p+f+1)\times(p+f+1)\) pour réduire le temps de calcul (annexe
\ref{sec-df-var}) : \[
\tr(\boldsymbol \Delta^2)=(n-(p+f))L_0^2+2\sum_{k=1}^{p+f}(n-(p+f)-k)L_k^2
\] où \(L_0,\dots,L_{p+f}\) sont définis par : \[
\begin{pmatrix}w_{-p} & \cdots & w_f\end{pmatrix}
\begin{pmatrix}
w_{-p} &0 & 0 &\cdots& 0 \\
w_{-p+1} & w_{-p} & 0 & \ddots & \vdots \\
w_{-p+2} & w_{-p+1}& w_{-p} & \ddots & \vdots\\
\vdots & \vdots & \vdots & \ddots &\vdots \\
w_f & w_{f-1} & w_{f-2}&\ddots&w_{-p}\\
\end{pmatrix}=\begin{pmatrix} L_0 & \cdots & L_{p+f} \end{pmatrix}
\] avec \(\boldsymbol w = \begin{pmatrix}w_{-p},\dots,w_f\end{pmatrix}\)
la moyenne mobile telle que \(w_0=1-\theta_0\) et \(w_i=-\theta_i\) pour
\(i\ne0\). C'est la méthode de calcul utilisée par défaut dans
\texttt{rjd3filters}. Le calcul de \(\nu\) peut aussi être approximé par
\(\tr(\boldsymbol \Delta)\)
(\texttt{rjd3filters::confint\_filter(exact\_df\ =\ FALSE)}) ce qui
permet de réduire davantage le temps de calcul\footnote{ Sur une série
  mensuelle de 19 ans (\(n=228\)), en utilisant la moyenne mobile de
  Henderson de 13 termes (\(p=f=6\)), le temps de calcul \(\nu\) est
  d'en moyenne 0,11 seconde (sur 1 000 évaluations). Il est divisé
  d'environ 350 en utilisant la formule proposée dans cet article
  (environ 0,31 milliseconde) et d'environ 5 800 en utilisant
  l'approximation \(\nu\simeq \tr(\boldsymbol \Delta)\) (environ 0,02
  milliseconde).}.

\begin{remark}
Lorsque l'estimation de la tendance-cycle est basée sur des moyennes
mobiles construites à l'aide d'un régresseur externe
(section~\ref{sec-constr-mm}), les formules ici présentées ne sont pas
applicables puisque des moyennes mobiles différentes sont utilisées pour
l'estimation finale. Pour la construction d'intervalles de confiance, le
plus simple est alors de reconstruire la matrice \(\boldsymbol H.\) Pour
les estimations récentes, lorsque des moyennes mobiles asymétriques sont
utilisées, le calcul des intervalles de confiance nécessite la création
d'une matrice \(\boldsymbol H\) fictive (qui correspondrait à la matrice
utilisée si toutes les estimations étaient faites avec une moyenne
mobile asymétrique). Toutefois, lorsque plusieurs outliers consécutifs
sont utilisés dans le passé, la construction d'une moyenne mobile
asymétrique robuste n'est pas toujours possible : dans ce cas il est
possible d'utiliser par défaut la moyenne mobile asymétrique linéaire de
référence (ce qui a pour effet d'augmenter la variance estimée). C'est
l'approche implémentée dans \texttt{rjd3filters}.
\end{remark}

\begin{remark}
La matrice \(\boldsymbol H\) pourrait également être construite en
complétant les \(p\) premières et \(f\) dernières lignes par les
moyennes mobiles asymétriques utilisées. Toutefois, puisque les modèles
sous-jacents aux moyennes mobiles asymétriques sont différents de ceux
des moyennes mobiles symétriques (par exemple Musgrave modélise une
tendance de degré 1 alors que Henderson modélise une tendance de degré
3), on peut supposer les variances différentes pour chaque moyenne
mobile et estimer celles associées aux moyennes mobiles asymétriques en
utilisant l'ensemble des données. C'est l'approche implémentée dans
\texttt{rjd3filters}.
\end{remark}

\section{Résultats}\label{sec-results}

Les différentes méthodes sont tout d'abord comparées sur des séries
simulées (choc de 10 \%, tendance de degré 0, 1 ou 2,
section~\ref{sec-res-series-simul}) puis sur des séries réelles
(section~\ref{sec-res-series-reelles}).

\subsection{Séries simulées}\label{sec-res-series-simul}

\begin{figure}[H]

\caption{\label{fig-simul-y}Séries simulées avec un choc ponctuel
(\emph{additive outlier}, AO) ou permanent (\emph{level shift}, LS) en
janvier 2022}

\centering{

\includegraphics{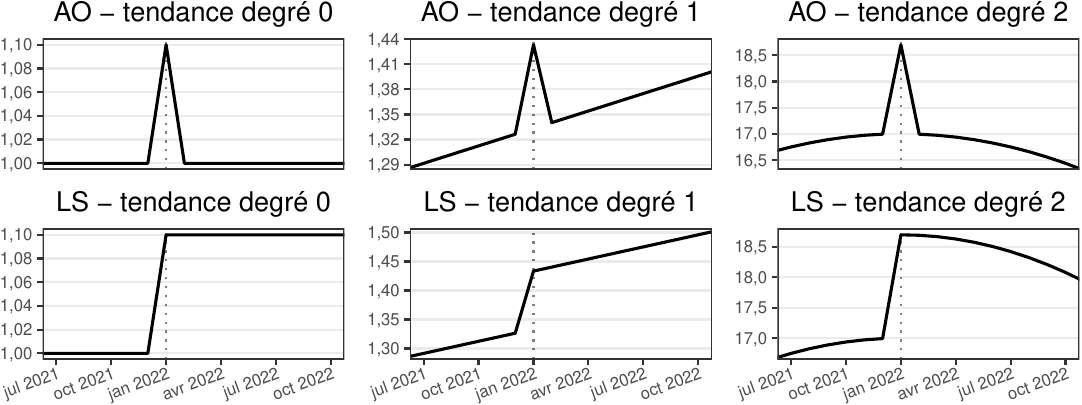}

}

\end{figure}

Afin d'illustrer les performances des différentes méthodes, les
estimations sont comparées sur des séries simulées dans un cas extrême :
lorsque l'irrégulier est nul. Trois tendances sont simulées, de degré 0,
1 ou 2, et un choc positif de 10 \% est introduit en janvier 2022 : soit
un choc ponctuel (\emph{additive outlier}, affectant l'irrégulier) soit
un choc permanent (\emph{level shift}, affectant la tendance-cycle). Les
séries simulées commencent en janvier 2018, soit 4 ans avant le choc
afin de limiter l'impact des points atypiques sur le calcul des
intervalles de confiance (puisque l'irrégulier est nul, l'intervalle de
confiance est également nul avant l'introduction du choc). La
figure~\ref{fig-simul-y} montre les différentes séries simulées.

\begin{figure}[H]

\caption{\label{fig-simul-ao-out-td0-est}Estimations en temps réel de la
tendance-cycle pour une série simulée avec une tendance de degré 0 et un
choc ponctuel (\emph{additive outlier}, AO) en janvier 2022}

\centering{

\includegraphics{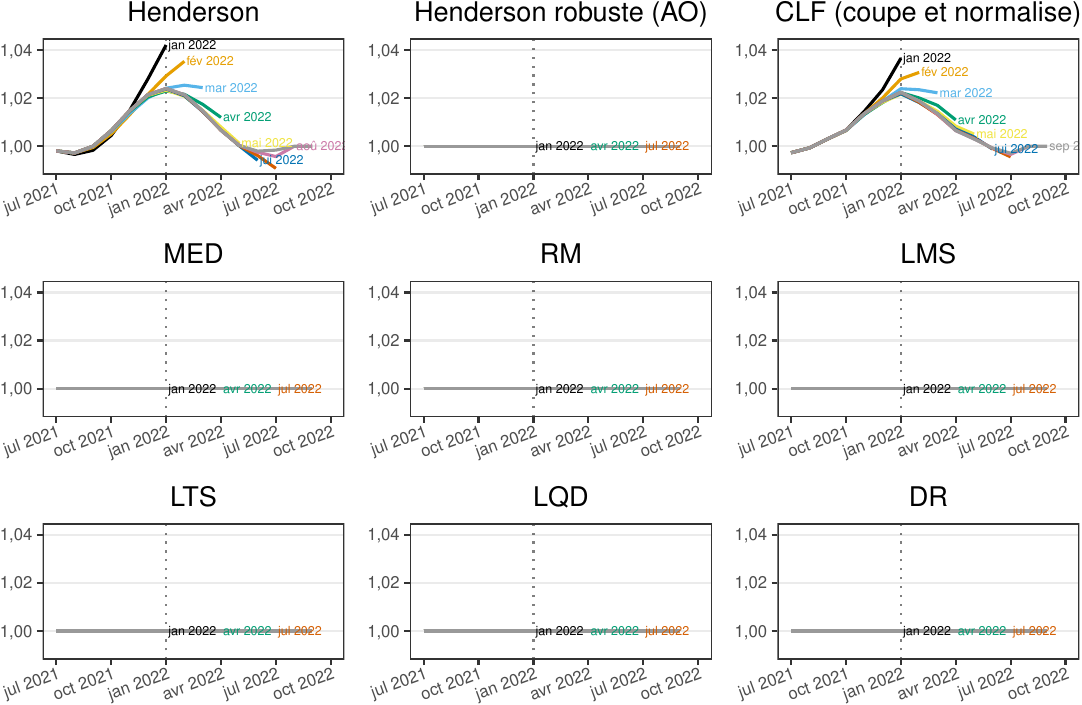}

\scriptsize\raggedright

\textbf{Note} : Henderson : Henderson (estimation finale) et Musgrave
(estimations intermédiaires) ; Henderson robuste : moyennes mobiles
robustes de Henderson (estimation finale) et de Musgrave (estimations
intermédiaires) ; CLF : \emph{cascade linear filter} (estimation finale)
et méthode « couper-et-normaliser » (estimations intermédiaires) ; MED :
Médiane mobile ; RM : médiane répétée ; LMS : moindres carrés médians ;
LTS : moindres carrés élagués ; LQD : moindres quartiles différenciés ;
DR : Régression profonde.

}

\end{figure}

\begin{figure}[H]

\caption{\label{fig-simul-ao-out-td0-ci}Intervalles de confiance pour
les filtres de Henderson et filtres de Henderson robustes robuste pour
une série simulée avec une tendance de degré 0 et un choc ponctuel
(\emph{additive outlier}, AO) en janvier 2022}

\centering{

\includegraphics{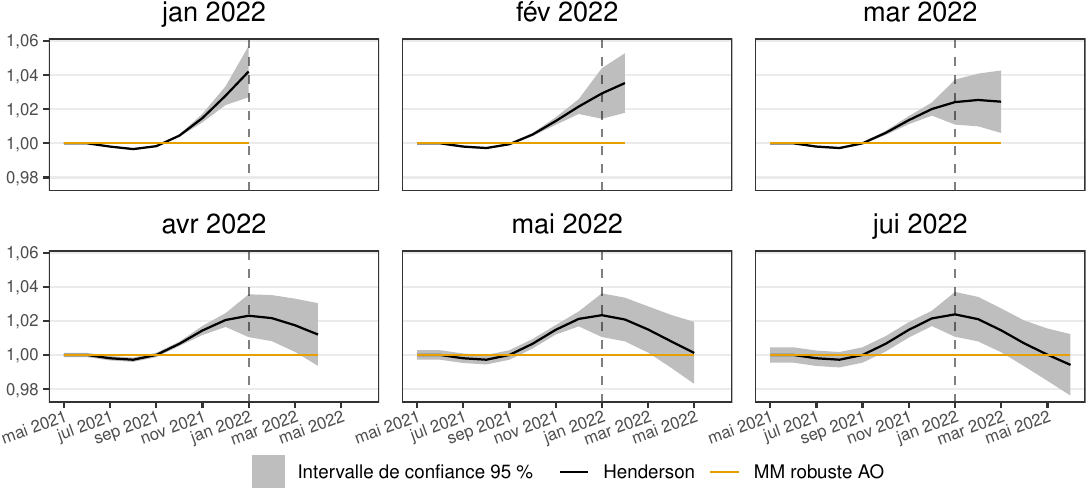}

}

\end{figure}

\begin{figure}[H]

\caption{\label{fig-simul-ao-out-td1-est}Estimations en temps réel de la
tendance-cycle pour une série simulée avec une tendance de degré 1 et un
choc ponctuel (\emph{additive outlier}, AO) en janvier 2022}

\centering{

\includegraphics{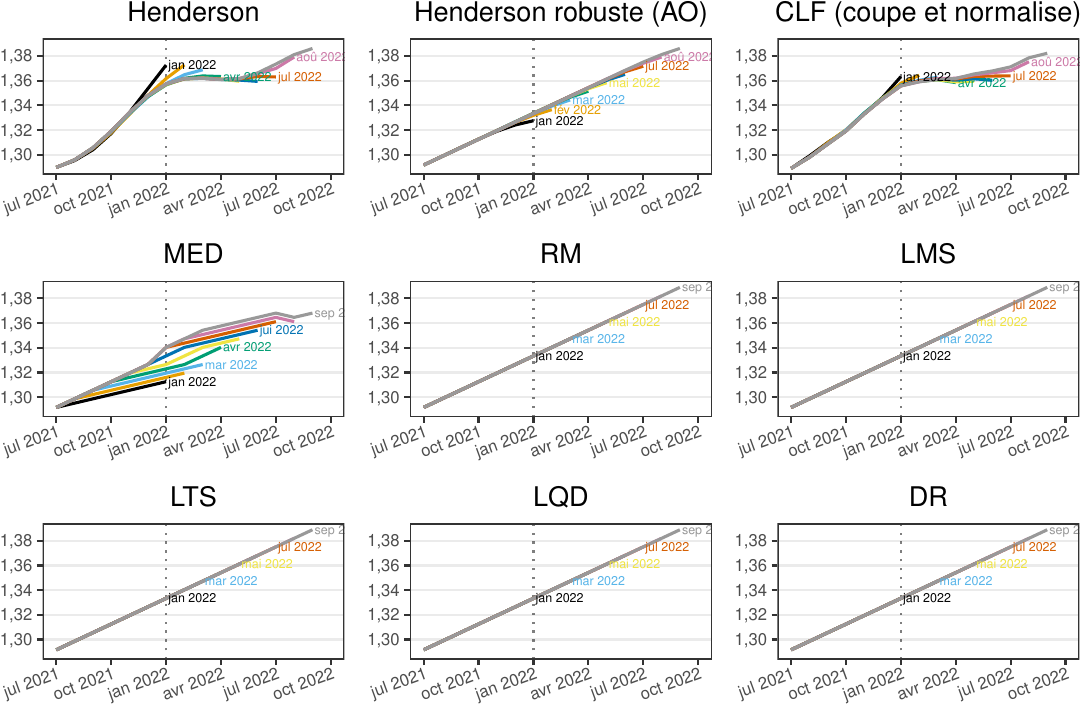}

\scriptsize\raggedright

\textbf{Note} : Henderson : Henderson (estimation finale) et Musgrave
(estimations intermédiaires) ; Henderson robuste : moyennes mobiles
robustes de Henderson (estimation finale) et de Musgrave (estimations
intermédiaires) ; CLF : \emph{cascade linear filter} (estimation finale)
et méthode « couper-et-normaliser » (estimations intermédiaires) ; MED :
Médiane mobile ; RM : médiane répétée ; LMS : moindres carrés médians ;
LTS : moindres carrés élagués ; LQD : moindres quartiles différenciés ;
DR : Régression profonde.

}

\end{figure}

\begin{figure}[H]

\caption{\label{fig-simul-ao-out-td1-ci}Intervalles de confiance pour
les filtres de Henderson et filtres de Henderson robustes pour une série
simulée avec une tendance de degré 1 et un choc ponctuel (\emph{additive
outlier}, AO) en janvier 2022}

\centering{

\includegraphics{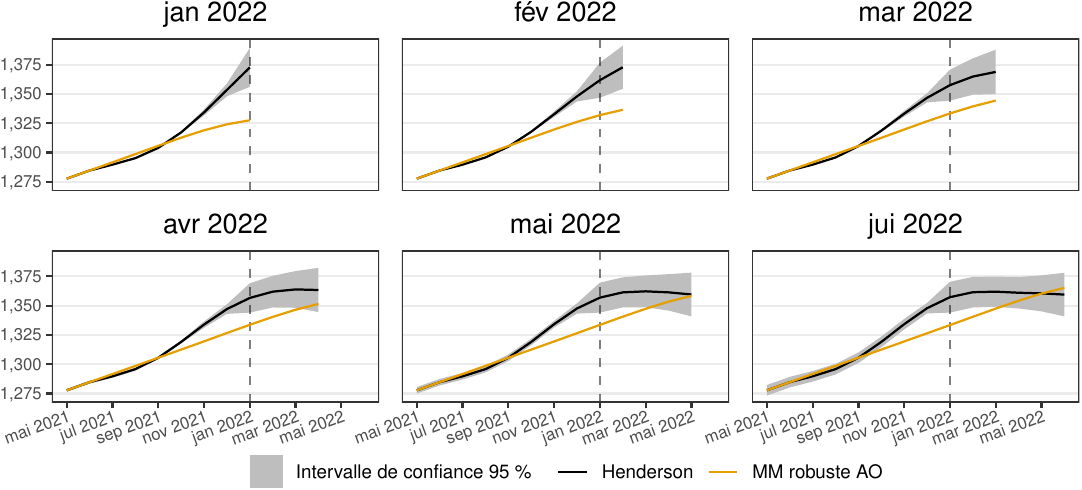}

}

\end{figure}

\begin{figure}[H]

\caption{\label{fig-simul-ao-out-td2-est}Estimations en temps réel de la
tendance-cycle pour une série simulée avec une tendance de degré 2 et un
choc ponctuel (\emph{additive outlier}, AO) en janvier 2022}

\centering{

\includegraphics{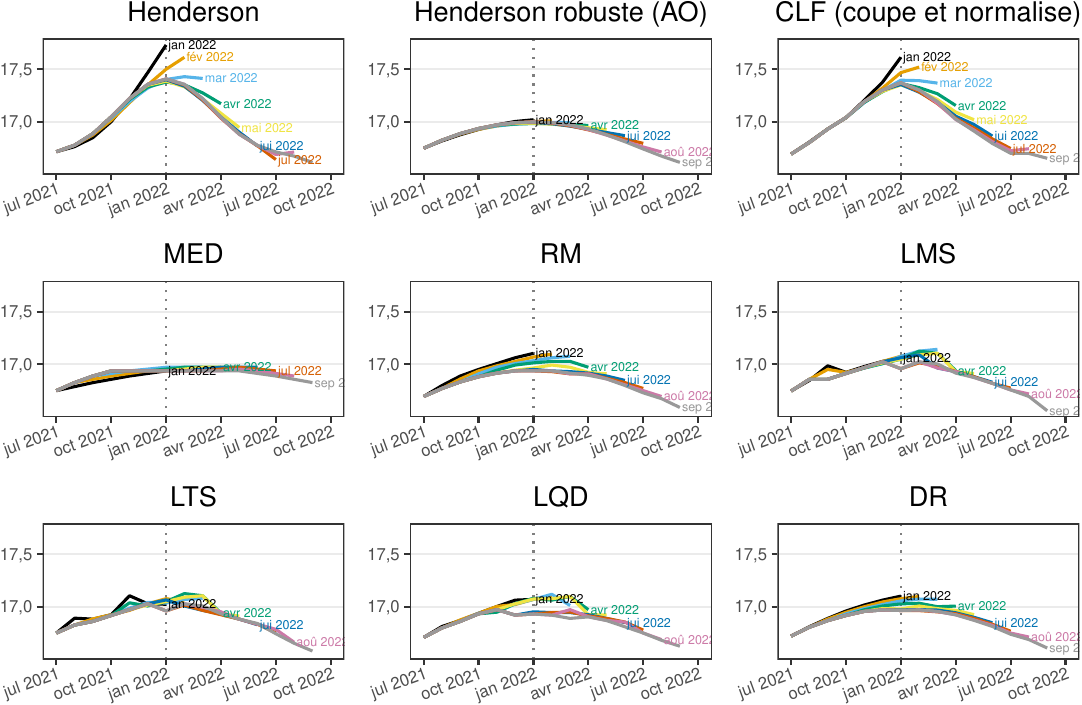}

\scriptsize\raggedright

\textbf{Note} : Henderson : Henderson (estimation finale) et Musgrave
(estimations intermédiaires) ; Henderson robuste : moyennes mobiles
robustes de Henderson (estimation finale) et de Musgrave (estimations
intermédiaires) ; CLF : \emph{cascade linear filter} (estimation finale)
et méthode « couper-et-normaliser » (estimations intermédiaires) ; MED :
Médiane mobile ; RM : médiane répétée ; LMS : moindres carrés médians ;
LTS : moindres carrés élagués ; LQD : moindres quartiles différenciés ;
DR : Régression profonde.

}

\end{figure}

\begin{figure}[H]

\caption{\label{fig-simul-ao-out-td2-ci}Intervalles de confiance pour
les filtres de Henderson et filtres de Henderson robustes robuste pour
une série simulée avec une tendance de degré 1 et un choc ponctuel
(\emph{additive outlier}, AO) en janvier 2022}

\centering{

\includegraphics{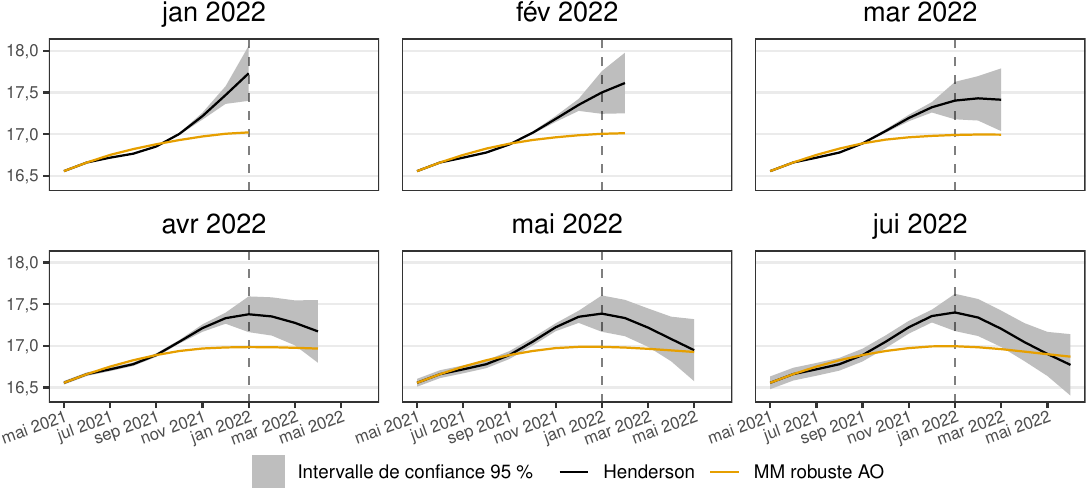}

}

\end{figure}

Pour les chocs ponctuels (\emph{additive outlier}, AO), lorsque la
tendance est de degré 0, aucune des méthodes robustes n'est influencée
par la présence de l'AO, tout comme la moyenne mobile de Henderson
robuste (figure~\ref{fig-simul-ao-out-td0-est}). Les estimations avec
les moyennes mobiles de Henderson et CLF sont biaisées avec des
révisions importantes à la date du choc, légèrement plus petites pour la
méthode CLF. L'analyse des intervalles de confiance valident clairement
la présence du point atypique (figure~\ref{fig-simul-ao-out-td0-ci}).\\
Lorsque la tendance est de degré 1 les résultats sont similaires sauf
pour la médiane mobile qui sous-estime la tendance pour les estimations
intermédiaires (figures \ref{fig-simul-ao-out-td1-est} et
\ref{fig-simul-ao-out-td1-ci}). Comme notamment indiqué par
\textcite{gather2006online}, cela suggère d'utiliser des quantiles plus
élevés que la médiane pour les estimations en temps réel avec les
méthodes robustes. Un léger biais négatif s'observe pour la méthode de
Henderson robuste : cela provient du fait que pour la création de ces
moyennes mobiles la pente est fixé (ratio \(\delta_1/\sigma\) fixé et
\(\sigma\) a priori fixé) à une valeur différente de la pente simulée.
Cela suggère de préférer une paramétrisation locale de ce paramètre,
comme proposé par \textcite{jos2024AQLT}\footnote{ C'est ce qui a été
  implémenté dans la fonction
  \texttt{publishTC::henderson\_robust\_smoothing(local\_icr\ =\ TRUE)}.}.\\
Lorsque la tendance est de degré 2, les résultats avec les moyennes
mobiles sont similaires (révisions importantes autour du choc pour
Henderson et CLF et faibles pour Henderson robuste, figures
\ref{fig-simul-ao-out-td2-est} et \ref{fig-simul-ao-out-td2-ci}). Pour
les méthodes robustes, les estimations finales reproduisent bien le
retournement de tendance malgré le fait qu'elles ne modélisent qu'une
tendance de degré 1. Pour les estimations en temps réel, les tendances
sont surestimées mais les révisions faibles : cela s'explique par le
faible degré de courbure de la série simulée et pourraient être
minimisées en modélisant des tendances de degré supérieur (voir annexe
\ref{sec-an-lms-lts} pour les résultats avec les méthodes LMS et LTS).

\begin{figure}[H]

\caption{\label{fig-simul-ls-out-td0-est}Estimations en temps réel de la
tendance-cycle pour une série simulée avec une tendance de degré 0 et un
choc permanent (\emph{level shift}, LS) en janvier 2022}

\centering{

\includegraphics{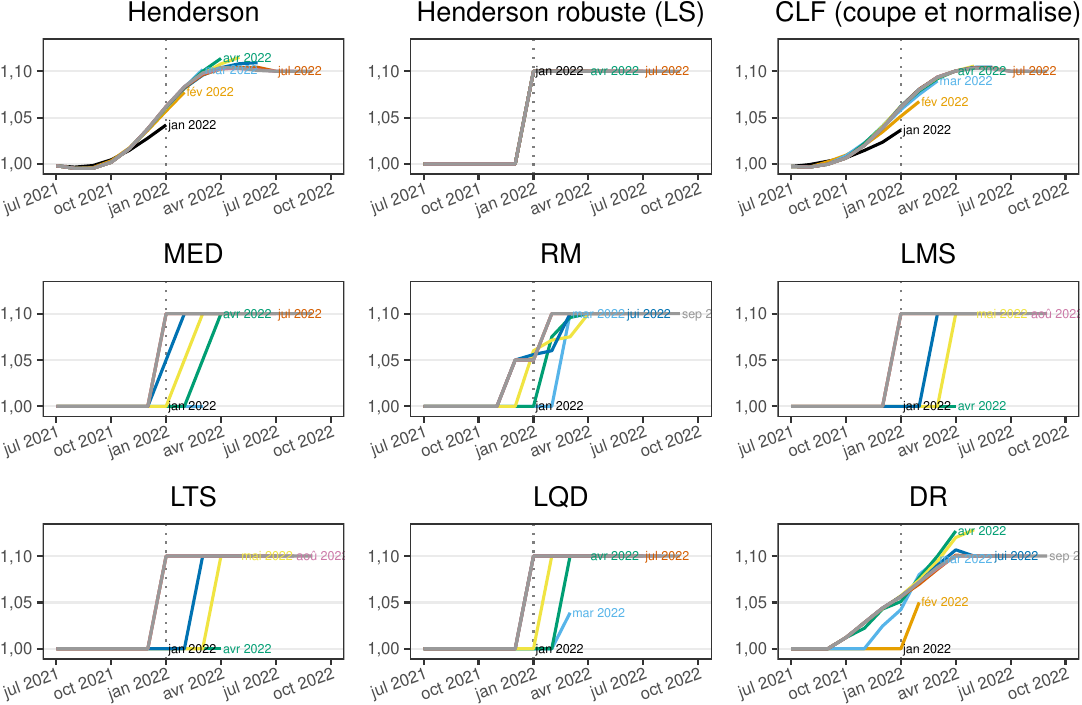}

\scriptsize\raggedright

\textbf{Note} : Henderson : Henderson (estimation finale) et Musgrave
(estimations intermédiaires) ; Henderson robuste : moyennes mobiles
robustes de Henderson (estimation finale) et de Musgrave (estimations
intermédiaires) ; CLF : \emph{cascade linear filter} (estimation finale)
et méthode « couper-et-normaliser » (estimations intermédiaires) ; MED :
Médiane mobile ; RM : médiane répétée ; LMS : moindres carrés médians ;
LTS : moindres carrés élagués ; LQD : moindres quartiles différenciés ;
DR : Régression profonde.

}

\end{figure}

\begin{figure}[H]

\caption{\label{fig-simul-ls-out-td0-ci}Intervalles de confiance pour
les filtres de Henderson et filtres de Henderson robustes robuste pour
une série simulée avec une tendance de degré 0 et un choc permanent
(\emph{level shift}, LS) en janvier 2022}

\centering{

\includegraphics{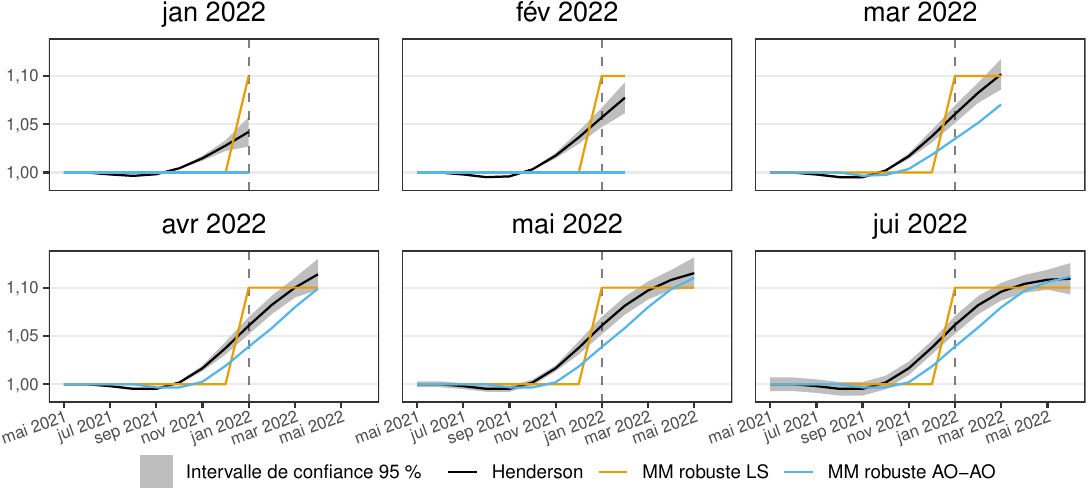}

}

\end{figure}

\begin{figure}[H]

\caption{\label{fig-simul-ls-out-td1-est}Estimations en temps réel de la
tendance-cycle pour une série simulée avec une tendance de degré 1 et un
choc permanent (\emph{level shift}, LS) en janvier 2022}

\centering{

\includegraphics{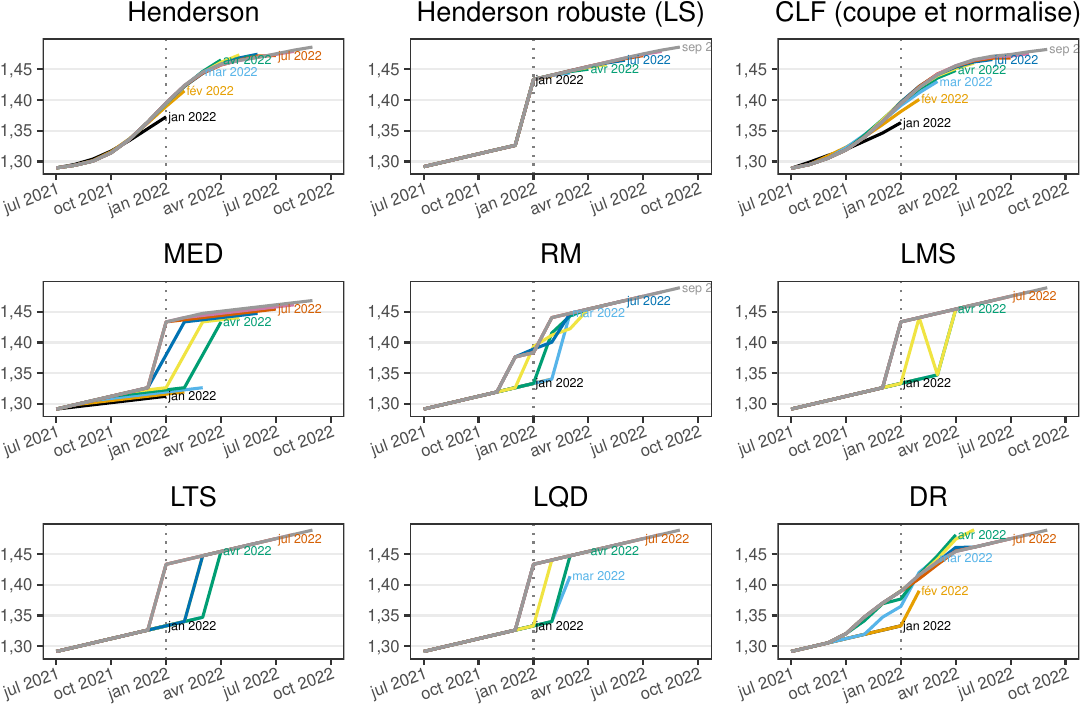}

\scriptsize\raggedright

\textbf{Note} : Henderson : Henderson (estimation finale) et Musgrave
(estimations intermédiaires) ; Henderson robuste : moyennes mobiles
robustes de Henderson (estimation finale) et de Musgrave (estimations
intermédiaires) ; CLF : \emph{cascade linear filter} (estimation finale)
et méthode « couper-et-normaliser » (estimations intermédiaires) ; MED :
Médiane mobile ; RM : médiane répétée ; LMS : moindres carrés médians ;
LTS : moindres carrés élagués ; LQD : moindres quartiles différenciés ;
DR : Régression profonde.

}

\end{figure}

\begin{figure}[H]

\caption{\label{fig-simul-ls-out-td1-ci}Intervalles de confiance pour
les filtres de Henderson et filtres de Henderson robustes robuste pour
une série simulée avec une tendance de degré 1 et un choc permanent
(\emph{level shift}, LS) en janvier 2022}

\centering{

\includegraphics{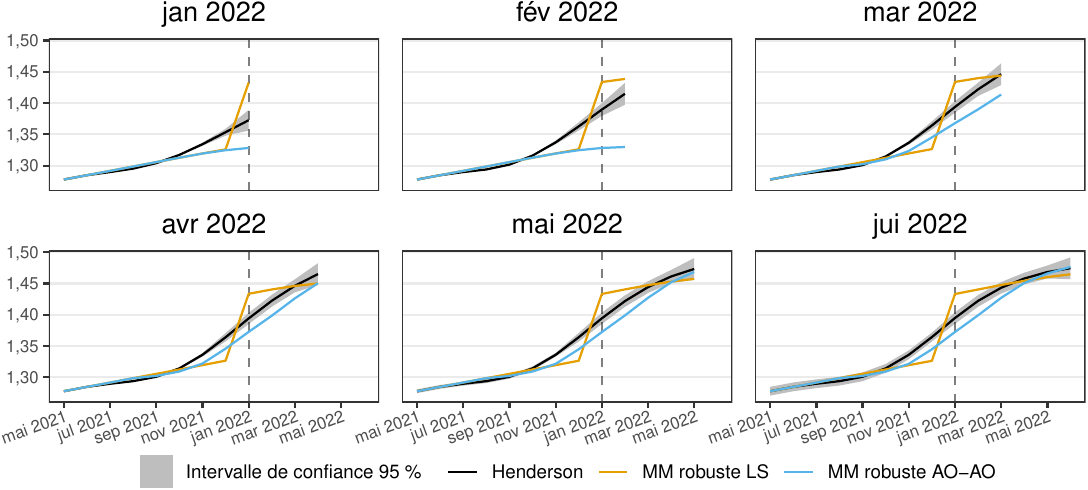}

}

\end{figure}

\begin{figure}[H]

\caption{\label{fig-simul-ls-out-td2-est}Estimations en temps réel de la
tendance-cycle pour une série simulée avec une tendance de degré 2 et un
choc permanent (\emph{level shift}, LS) en janvier 2022}

\centering{

\includegraphics{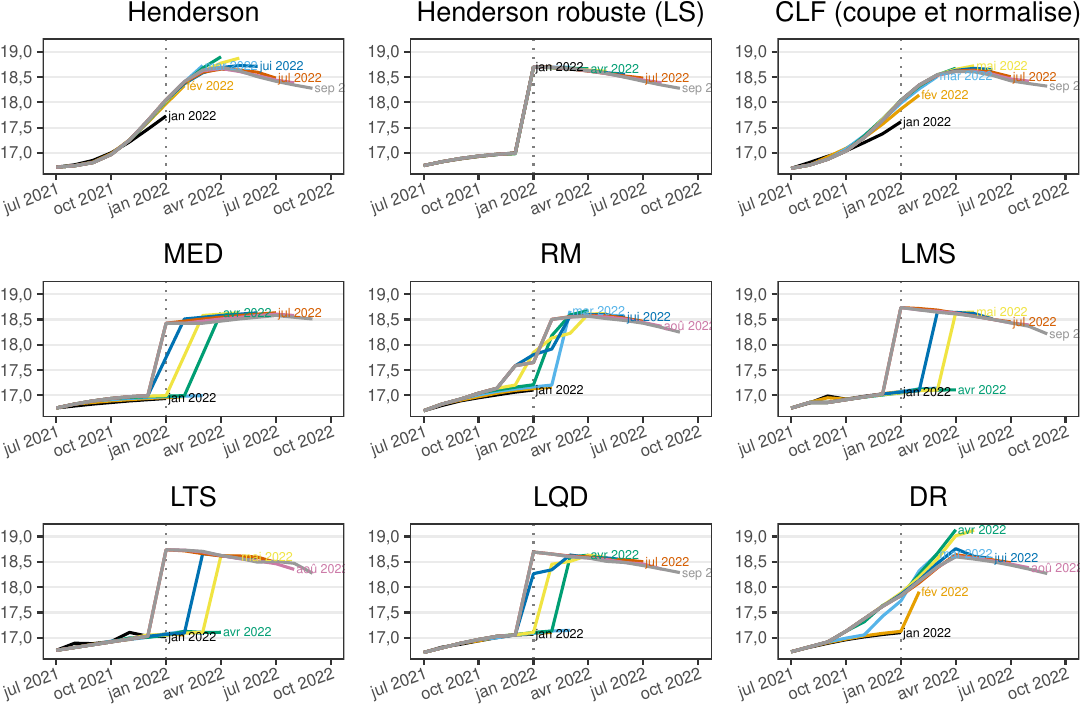}

\scriptsize\raggedright

\textbf{Note} : Henderson : Henderson (estimation finale) et Musgrave
(estimations intermédiaires) ; Henderson robuste : moyennes mobiles
robustes de Henderson (estimation finale) et de Musgrave (estimations
intermédiaires) ; CLF : \emph{cascade linear filter} (estimation finale)
et méthode « couper-et-normaliser » (estimations intermédiaires) ; MED :
Médiane mobile ; RM : médiane répétée ; LMS : moindres carrés médians ;
LTS : moindres carrés élagués ; LQD : moindres quartiles différenciés ;
DR : Régression profonde.

}

\end{figure}

\begin{figure}[H]

\caption{\label{fig-simul-ls-out-td2-ci}Intervalles de confiance pour
les filtres de Henderson et filtres de Henderson robustes robuste pour
une série simulée avec une tendance de degré 2 et un choc permanent
(\emph{level shift}, LS) en janvier 2022}

\centering{

\includegraphics{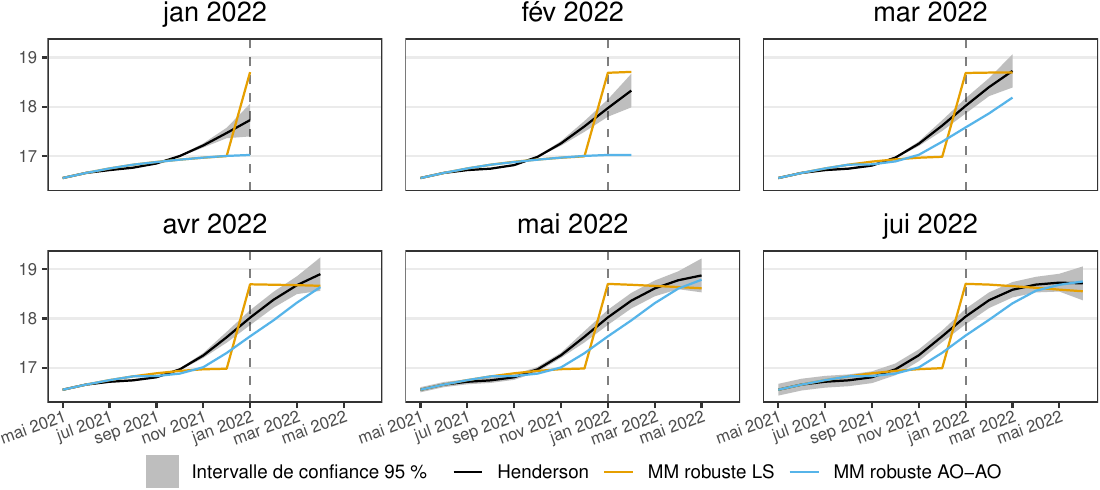}

}

\end{figure}

Pour les chocs permanents (\emph{level shift}, LS), à l'exception de la
régression profonde (DR) et de la régression médiane répétée (RM), les
estimations finales de la tendance-cycle avec les méthodes robustes ne
sont pas influencées par la présence du choc (figures
\ref{fig-simul-ls-out-td0-est}, \ref{fig-simul-ls-out-td1-est} et
\ref{fig-simul-ls-out-td2-est}). En revanche, aucune des estimations
intermédiaires ne prend en compte le changement de niveau à la bonne
date : il faut attendre l'estimation de juin 2022 pour que le changement
de niveau soit pris en compte à la bonne date pour la méthode LQD et
juillet 2022 (soit l'estimation finale) pour les autres méthodes, même
lorsque la tendance est de degré 0. Modéliser, dans ces méthodes, une
tendance de degré 2 ne change pas ce résultat (annexe
\ref{sec-an-lms-lts}). Cela suggère encore d'utiliser des quantiles plus
élevés que la médiane pour les estimations en temps réel.\\
Les moyennes mobiles linéaires classiques (Henderson et CLF) donnent des
résultats similaires~: la rupture en niveau est lissée, les points avant
le choc sont donc sur-estimés et ceux après le choc sous-estimés. Il y a
également d'importantes révisions pour la première estimation de janvier
2022 (date du choc). Il n'y a en revanche quasiment aucune révision pour
les moyennes mobile de Henderson robuste à un choc permanent en janvier
2022.\\
Les intervalles de confiance des estimations de la moyenne mobile de
Henderson sont comparés à la moyenne mobile de Henderson robuste à un
choc permanent en janvier 2022 ou à un choc ponctuel en janvier et
février 2022 (figures \ref{fig-simul-ls-out-td0-ci},
\ref{fig-simul-ls-out-td1-ci} et \ref{fig-simul-ls-out-td2-ci}). Cela
permet de simuler le cas où on ne sait pas, en temps réel, trancher sur
la nature du choc : soit un choc permanent, soit un choc ponctuel
(lorsque l'on se place en janvier 2022), soit deux chocs ponctuels
(lorsque l'on se place en février 2022). En mars 2022, l'observation de
trois chocs de même ampleur confirme qu'il s'agit d'un choc permanent et
non ponctuel. Si l'on s'est trompé dans la nature du choc, jusqu'en
février 2022 les révisions sont nulles pour les estimations avant le
choc mais importantes pour celles après (janvier et février, puisque le
choc est affecté à l'irrégulier et non pas à la tendance-cycle).

\subsection{Séries réelles}\label{sec-res-series-reelles}

Puisque les résultats sont similaires à ceux sur les données simulées,
les performances des différentes méthodes sur des séries réelles lors
d'un choc ponctuel (AO) et d'un choc permanent (LS) sont données dans
l'annexe \ref{sec-autres-ex} : les moyennes mobiles classiques
(Henderson/Musgrave et CLF) donnent de fortes révisions autour des chocs
et ces révisions sont faibles avec les moyennes mobiles robustes.
Contrairement aux données simulées, sur les données réelles, même
lorsque la tendance semble être linéaire, les estimations intermédiaires
issues des méthodes robustes peuvent être très erratiques, conduisant à
des estimations parfois peu plausibles et des révisions importantes.

Dans cette section, nous nous concentrerons sur l'illustration d'autres
cas :

\begin{itemize}
\item
  Un choc ponctuel suivi d'un choc permanent
  (section~\ref{sec-ex-ao-ls}) à partir de l'analyse des
  immatriculations de voitures particulières neuves, corrigée des
  variations saisonnières (CVS) et des jours ouvrables (CJO), publiée
  par l'Insee (série
  \href{https://www.insee.fr/fr/statistiques/serie/010756763}{\texttt{010756763}}
  téléchargée en octobre 2024). Le choc ponctuel s'observe en août 2018,
  suivi d'un choc permanent en septembre 2018.
\item
  La crise financière de 2008 (section~\ref{sec-ex-crise-financiere}) à
  partir des ventes au détail et services de restauration aux
  États-Unis. Les données issues de la base FRED-MD \autocite{fredmd}
  contenant des séries économiques sur les États-Unis\footnote{ Les
    séries étudiées correspondent à la base publiée en novembre 2022.}
  (série \texttt{RETAILx}). Pour cette série la crise financière de 2008
  peut s'étudier comme deux chocs permanents consécutifs en octobre et
  novembre 2008.
\item
  La crise du COVID-19 (section~\ref{sec-ex-covid}) à partir l'indice de
  la production industrielle (IPI) dans l'industrie manufacturière
  CVS-CJO, publié par l'Insee (série
  \href{https://www.insee.fr/fr/statistiques/serie/010768307}{\texttt{010768307}}
  publiée le 04 octobre 2024). Deux autres exemples sont donnés dans
  l'annexe \ref{sec-autres-ex}.
\item
  Un point de retournement qui n'est pas associé à un choc
  (section~\ref{sec-ex-retournement}) à partir du niveau d'emploi aux
  États-Unis (série \texttt{CE16OV} de la base FRED-MD). Un autre
  exemple est donné dans l'annexe \ref{sec-autres-ex}.
\end{itemize}

\begin{figure}[H]

\caption{\label{fig-ex-y}Séries des immatriculations de voitures
particulières neuves en France, ventes au détail et services de
restauration aux États-Unis, IPI dans l'industrie manufacturière en
France et niveau d'emploi aux États-Unis}

\centering{

\includegraphics{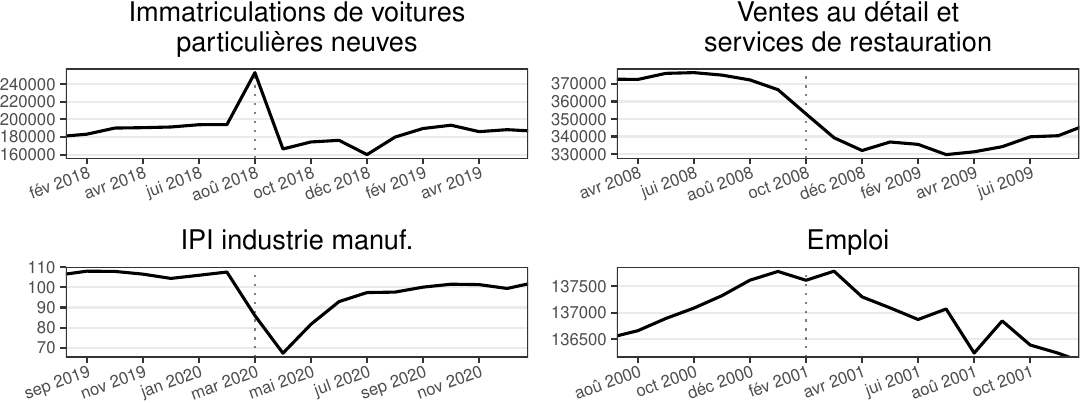}

}

\end{figure}

Toutes ces séries sont représentées dans la figure~\ref{fig-ex-y}.

\subsubsection{Choc ponctuel (AO) suivi d'un choc permanent
(LS)}\label{sec-ex-ao-ls}

L'entrée en vigueur de la procédure d'essai mondiale harmonisée pour les
véhicules légers (\emph{Worldwide harmonized Light vehicles Test
Procedures}, WLTP) en septembre 2018 a eu pour effet une forte hausse
des immatriculations de voitures particulières neuves en août 2018 (choc
ponctuel), les constructeurs ayant déstocké des modèles qui n'étaient
plus autorisés à la vente à partir de cette date. À partir du mois de
septembre, le niveau des immatriculations était plus faible par
contre-coup de la hausse du mois d'août. Même si ce choc en niveau
semble temporaire sur 5 mois (retour au niveau de juillet 2018 à partir
de février), il est, par simplification, modélisé dans la construction
des moyennes mobiles robustes comme un choc permanent.

\begin{figure}[H]

\caption{\label{fig-imat2018-est}Estimations en temps réel de la
tendance-cycle des immatriculations de voitures particulières neuves à
partir d'août 2018}

\centering{

\includegraphics{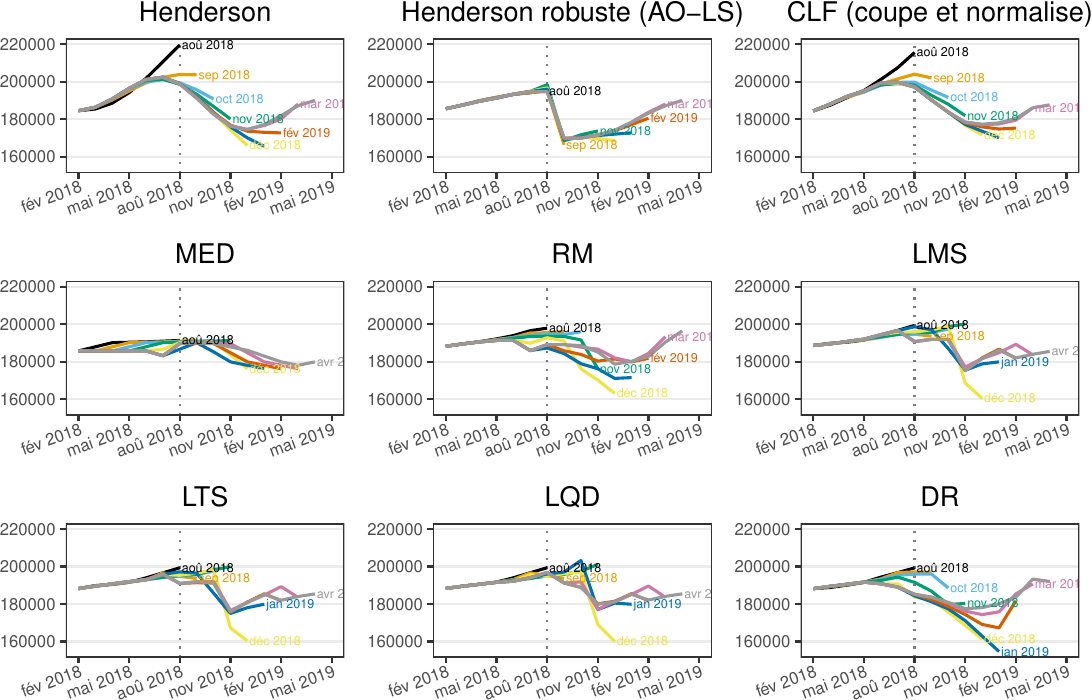}

\scriptsize\raggedright

\textbf{Note} : Henderson : Henderson (estimation finale) et Musgrave
(estimations intermédiaires) ; Henderson robuste : moyennes mobiles
robustes de Henderson (estimation finale) et de Musgrave (estimations
intermédiaires) ; CLF : \emph{cascade linear filter} (estimation finale)
et méthode « couper-et-normaliser » (estimations intermédiaires) ; MED :
Médiane mobile ; RM : médiane répétée ; LMS : moindres carrés médians ;
LTS : moindres carrés élagués ; LQD : moindres quartiles différenciés ;
DR : Régression profonde.

}

\end{figure}

\begin{figure}[H]

\caption{\label{fig-imat2018-ci}Intervalles de confiance pour les
filtres de Henderson et filtres de Henderson robustes robuste pour les
immatriculations de voitures particulières neuves à partir d'août 2018}

\centering{

\includegraphics{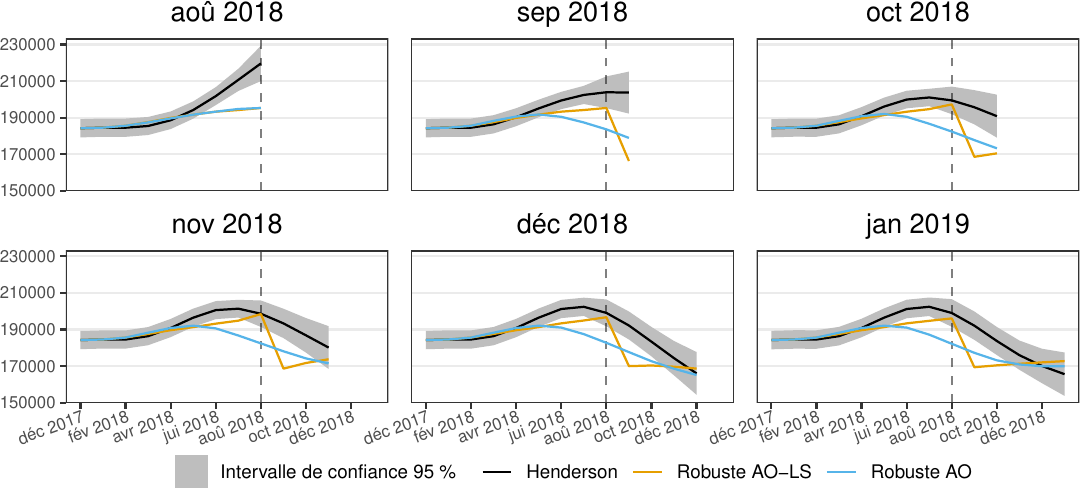}

}

\end{figure}

Les méthodes robustes ne sont pas affectées par le choc ponctuel mais ne
reproduisent pas le changement de niveau à la bonne date
(figure~\ref{fig-imat2018-est}). Les moyennes mobiles linéaires de
Henderson et CLF ont des résultats similaires : fortes révisions en août
2018, un point de retournement détecté en avance (juillet 2018). Les
moyennes mobiles robustes conduisent à peu de révisions autour des
chocs. Lorsque l'on ne modélise que le choc ponctuel d'août
(figure~\ref{fig-imat2018-ci}) un point de retournement est détecté en
mars 2018, ce qui parait peu plausible.

\subsubsection{Crise financière de 2008 (deux chocs permanents
consécutifs)}\label{sec-ex-crise-financiere}

\begin{figure}[H]

\caption{\label{fig-retailx2008-est}Estimations en temps réel de la
tendance-cycle des ventes au détail et services de restauration aux
États-Unis à partir d'octobre 2008}

\centering{

\includegraphics{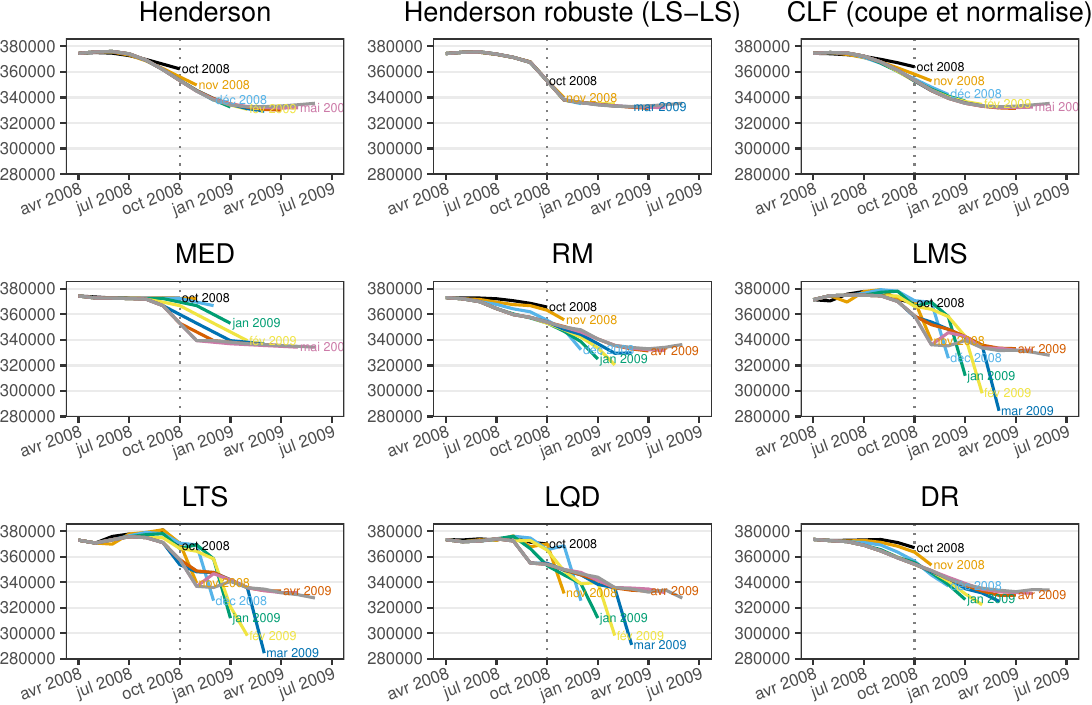}

\scriptsize\raggedright

\textbf{Note} : Henderson : Henderson (estimation finale) et Musgrave
(estimations intermédiaires) ; Henderson robuste : moyennes mobiles
robustes de Henderson (estimation finale) et de Musgrave (estimations
intermédiaires) ; CLF : \emph{cascade linear filter} (estimation finale)
et méthode « couper-et-normaliser » (estimations intermédiaires) ; MED :
Médiane mobile ; RM : médiane répétée ; LMS : moindres carrés médians ;
LTS : moindres carrés élagués ; LQD : moindres quartiles différenciés ;
DR : Régression profonde.

}

\end{figure}

\begin{figure}[H]

\caption{\label{fig-retailx2008-ci}Intervalles de confiance pour les
filtres de Henderson et filtres de Henderson robustes robuste pour les
ventes au détail et services de restauration aux États-Unis à partir
d'octobre 2008}

\centering{

\includegraphics{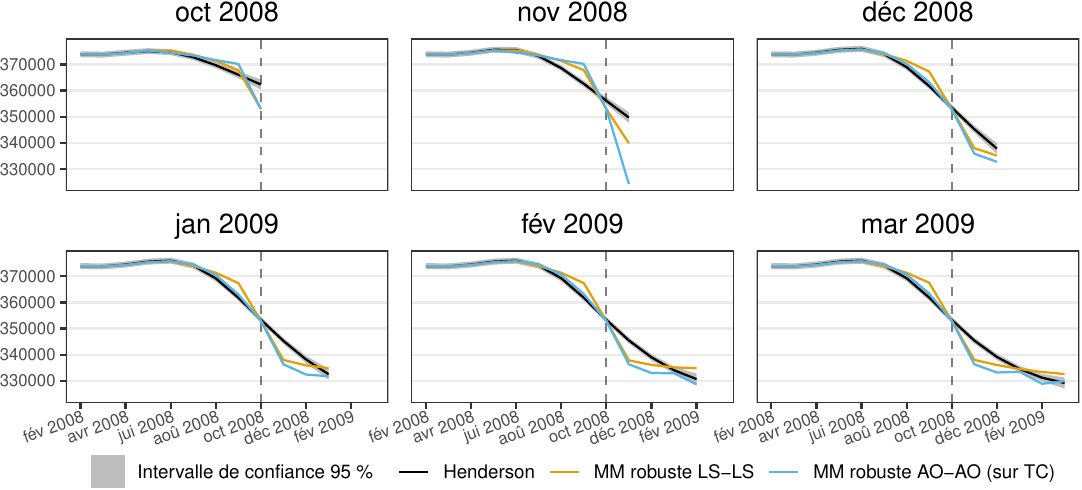}

}

\end{figure}

La crise financière de 2008 est illustrée à partir des ventes au détail
et services de restauration. Sur cette série, on observe deux chocs
consécutifs en niveau : en octobre et en novembre 2008. Seules deux
méthodes robustes (médiane mobile et LMS) reproduisent ces chocs dans
les estimations finales (figure~\ref{fig-retailx2008-est}) ; certaines
estimations intermédiaires sont éloignées des estimations finales (par
exemple pour les médianes mobiles ou les points de décembre 2008 et
janvier 2009 RM, LMS, LTS et LQD) ce qui conduit à des révisions
importantes. Les moyennes mobiles linéaires donnent des estimations avec
relativement peu de révisions (en particulier pour les moyennes mobiles
robustes) mais le retournement conjoncturel est détecté trop tôt pour
Henderson et CLF. En temps-réel, si l'on ne sait pas que ce sont des
chocs permanents, les chocs peuvent être modélisés comme deux chocs
ponctuels dont on attribuerait l'effet à la tendance-cycle. Cette
modélisation conduit à des révisions importantes entre novembre et
décembre 2008 (figure~\ref{fig-retailx2008-ci}). En mars 2009 (où l'on
se doute que les chocs sont permanents et non ponctuels), l'utilisation
de cette mauvaise spécifications donnent des estimations de la
tendance-cycle proches des moyennes mobiles de Henderson avant octobre
2008 (avec un point de retournement qui est donc détecté trop tôt) et
proches des moyennes mobiles robustes à deux chocs permanents après
octobre 2008.

\subsubsection{COVID-19}\label{sec-ex-covid}

\begin{figure}[H]

\caption{\label{fig-ipi-manuf-covid-est}Estimations en temps réel de la
tendance-cycle de l'IPI CVS-CJO dans l'industrie manufacturière à partir
de mars 2020}

\centering{

\includegraphics{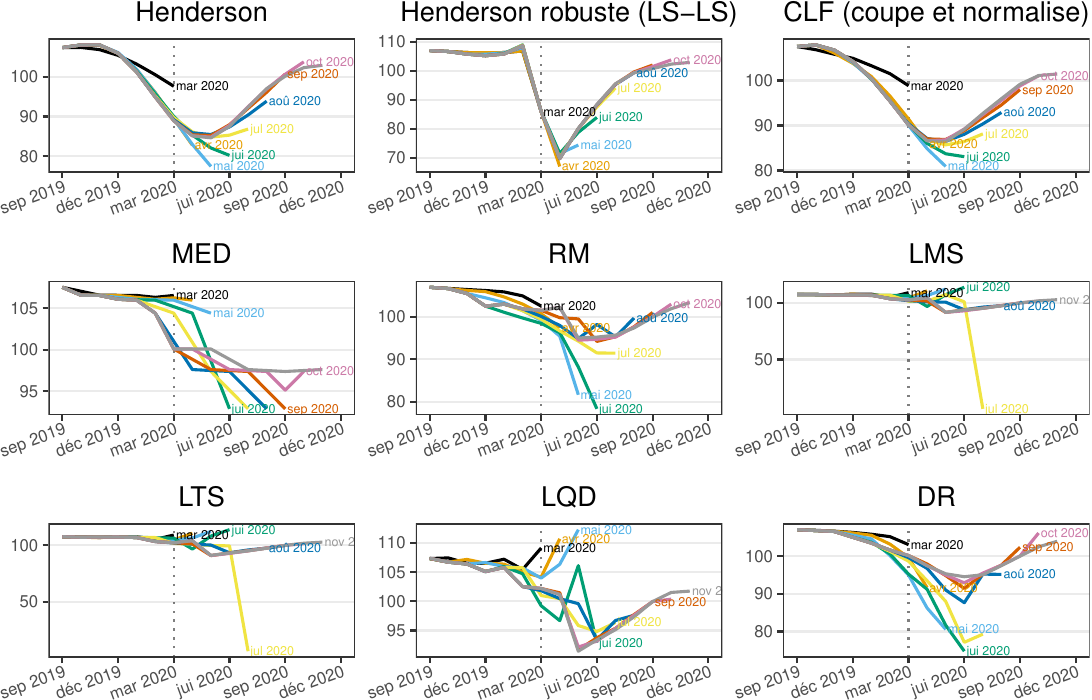}

\scriptsize\raggedright

\textbf{Note} : Henderson : Henderson (estimation finale) et Musgrave
(estimations intermédiaires) ; Henderson robuste : moyennes mobiles
robustes de Henderson (estimation finale) et de Musgrave (estimations
intermédiaires) ; CLF : \emph{cascade linear filter} (estimation finale)
et méthode « couper-et-normaliser » (estimations intermédiaires) ; MED :
Médiane mobile ; RM : médiane répétée ; LMS : moindres carrés médians ;
LTS : moindres carrés élagués ; LQD : moindres quartiles différenciés ;
DR : Régression profonde.

}

\end{figure}

Le COVID-19 est illustré sur l'indice de la production industrielle
(IPI) CVS-CJO dans l'industrie manufacturière. Sur cette série, on
observe deux chocs négatifs en mars et avril 2020, suivis d'une remontée
progressive jusqu'en juillet 2020 vers un niveau légèrement inférieur à
celui d'avant crise. Pour cette période, il est attendu de l'estimation
finale de la tendance-cycle qu'elle permette de détecter les deux points
de retournement en février 2020 (pic) et avril 2020 (creux). Les
méthodes robustes peinent à reproduire ces retournements : les
estimations finales sont robustes aux chocs, les révisions sont
importantes et les retournement conjoncturels détectés semblent plutôt
provenir des périodes post-chocs, ce qui conduit à détecter les points
de retournement avec décalage (figure~\ref{fig-ipi-manuf-covid-est}) Les
moyennes mobiles de Henderson et CLF conduisent, encore une fois, à
détecter les points de retournement avec un décalage ; les révisions
sont importantes autour des chocs (mars, mai et juin) et plus faibles
pour les autres mois.

\begin{figure}[H]

\caption{\label{fig-ipi-manuf-covid-ci}Intervalles de confiance pour les
filtres de Henderson et filtres de Henderson robustes robuste pour l'IPI
dans l'industrie manufacturière à partir de mars 2020}

\centering{

\includegraphics{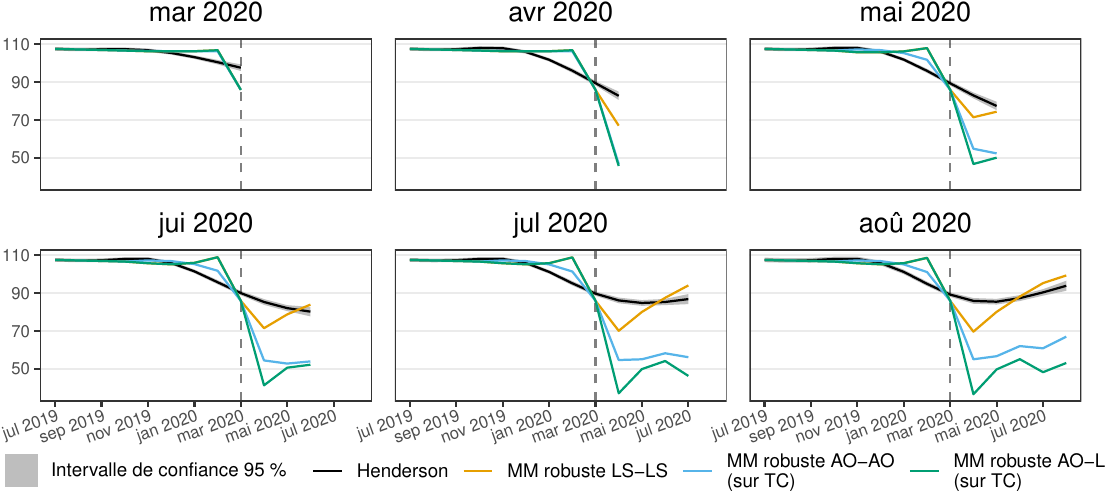}

}

\end{figure}

Pour la construction des moyennes mobiles robustes, même si l'on
pourrait construire des régresseurs spécifiques, il a ici été choisi de
rester sur la modélisation de chocs ponctuels et permanents ce qui
permet de garder une approche simple et plus facilement généralisable à
d'autres cas. La figure~\ref{fig-ipi-manuf-covid-ci} compare les
estimations des moyennes mobiles robustes construites en modélisant deux
chocs ponctuels consécutifs (dont on attribuerait l'effet à la
tendance-cycle plutôt qu'à l'irrégulier), deux chocs permanents
consécutifs ou un choc ponctuel suivi d'un choc permanent. La
modélisation de deux chocs permanents consécutifs en mars et avril 2020
permet de détecter les points de retournement aux bonnes dates, la
reprise de la production en mai est détecté dès la première estimation.
Ce n'est pas le cas de la modélisation par deux chocs ponctuels
consécutifs : il faut attendre les estimations de juin 2020 pour
reproduire le creux d'avril (les estimations de mai et juin donne une
baisse entre avril et mai) et le pic de février n'est pas conservé. La
modélisation d'un choc ponctuel suivi d'un choc permanent permet bien de
tracer le pic d'avril et le creux d'avril mais un pic indésirable est
créé en juin 2020. C'est donc la modélisation par deux chocs permanents
consécutifs qui donne ici les meilleurs résultats.

D'autres exemples sont donnés dans l'annexe \ref{sec-autres-ex-covid}.

\subsubsection{Points de retournement}\label{sec-ex-retournement}

\begin{figure}[H]

\caption{\label{fig-ce16ov2001-est}Estimations en temps réel de la
tendance-cycle de l'emploi aux États-Unis à partir de février 2001}

\centering{

\includegraphics{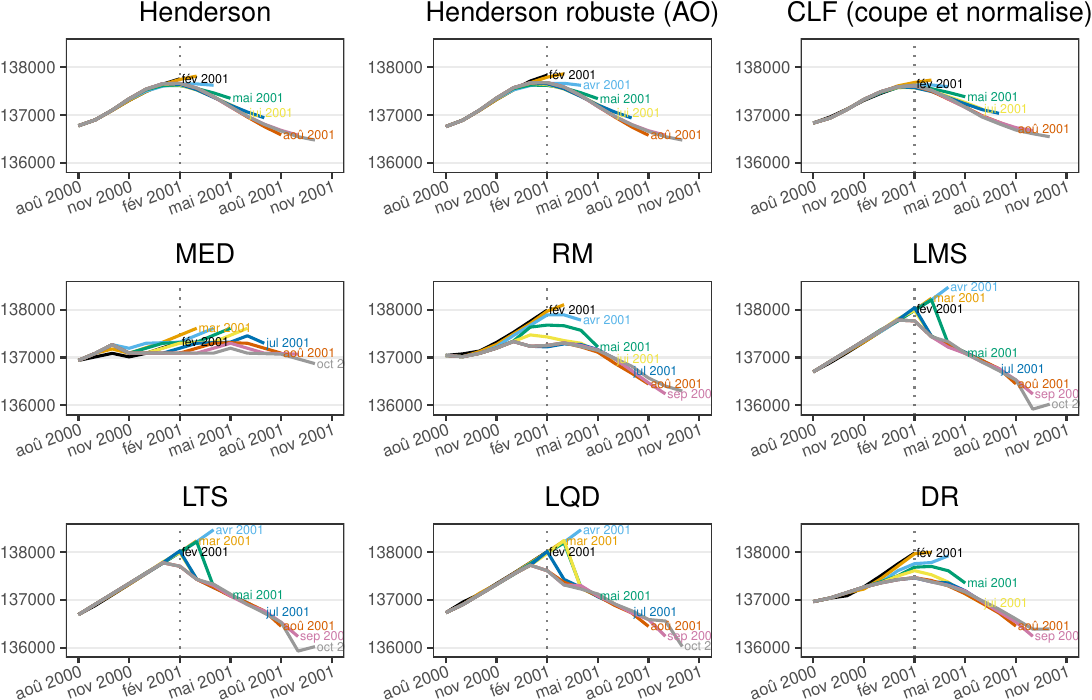}

\scriptsize\raggedright

\textbf{Note} : Henderson : Henderson (estimation finale) et Musgrave
(estimations intermédiaires) ; Henderson robuste : moyennes mobiles
robustes de Henderson (estimation finale) et de Musgrave (estimations
intermédiaires) ; CLF : \emph{cascade linear filter} (estimation finale)
et méthode « couper-et-normaliser » (estimations intermédiaires) ; MED :
Médiane mobile ; RM : médiane répétée ; LMS : moindres carrés médians ;
LTS : moindres carrés élagués ; LQD : moindres quartiles différenciés ;
DR : Régression profonde.

}

\end{figure}

\begin{figure}[H]

\caption{\label{fig-ce16ov2001-ci}Intervalles de confiance pour les
filtres de Henderson et filtres de Henderson robustes robuste pour
l'emploi aux États-Unis à partir de février 2001}

\centering{

\includegraphics{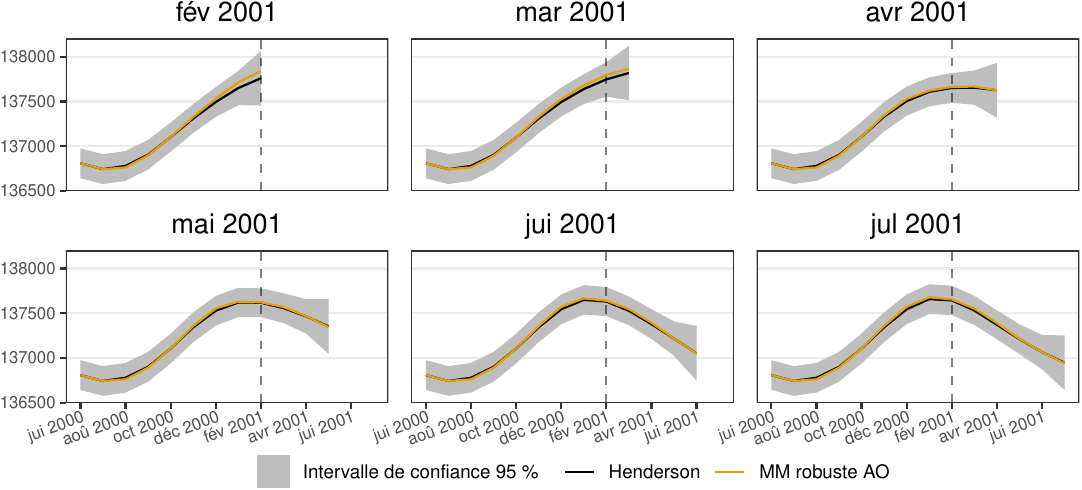}

}

\end{figure}

Un point de retournement, non associé à un choc, est illustré à partir
du niveau d'emploi aux États-Unis autour de février 2001. Parmi les
méthodes robustes, seule la régression profonde (DR) permet de détecter
le point de retournement à la bonne date (à partir des estimations de
juin 2021, figure~\ref{fig-ce16ov2001-est}) : les médianes mobiles (MED)
et les médianes répétées (RM) ne détectent aucun point de retournement
et les autres méthodes le détecte en janvier 2001 (à partir des
estimations de juillet 2001, pour les estimations antérieures il est
bien détecté en février). En revanche, modéliser une tendance locale de
degré 2 permet de détecter le point de retournement à la bonne date
(annexe \ref{sec-an-lms-lts}) Les révisions sont plus faibles pour les
estimations issues de moyennes mobiles et le fait de considérer à tort
le mois de février comme un choc ponctuel n'a pas d'impact sur les
estimations. Les estimations issues des moyennes mobiles robustes
restent dans l'intervalle de confiance des estimations des moyennes
mobiles de Henderson (figure~\ref{fig-ce16ov2001-ci}).

\section{Conclusion}\label{conclusion}

En conclusion, cette étude montre comment, à partir des moyennes mobiles
de Henderson et de Musgrave (répandues pour l'extraction de la
tendance-cycle) il est possible d'ajouter des régresseurs externe pour
construire des moyennes mobiles avec une meilleure décomposition
tendance-cycle et irrégulier. Cela permet notamment de modéliser des
chocs ponctuels (dont l'effet est généralement associé à l'irrégulier)
et des chocs permanents (dont l'effet est associé à la tendance-cycle).

Les moyennes mobiles linéaires classiques (Henderson, Musgrave,
\emph{cascade linear filter}\ldots) vont naturellement conduire à des
estimations biaisées : lors d'un choc permanent positif, la rupture va
être lissée et la tendance-cycle va être sur-estimée avant le choc et
sous-estimée après le choc~; lors d'un choc ponctuel (dont l'effet est
attribué à l'irrégulier), la tendance-cycle va être sur ou sous-estimée
autour du choc (en fonction du signe du choc). Cela va également amener
à des fortes révisions des estimations autour des chocs. Lorsqu'un choc
est associé à un point de retournement de l'économie (par exemple lors
de la crise financière de 2008 ou lors de la pandémie du COVID-19), les
estimations de la tendance-cycle vont conduire à un décalage dans la
détection des points de retournement (contraction détectée en avance et
reprise en retard), ce qui peut perturber l'utilisateur lors de
l'analyse de ces séries. À l'inverse, la construction de moyennes
mobiles robustes permettent de minimiser les révisions des chocs et de
bien détecter les points de retournement. La comparaison entre moyennes
mobiles robustes et classiques, à l'aide de la construction
d'intervalles de confiance et d'une analyse économique, permet de
valider la modélisation retenue. De plus, lorsque aucun choc n'est
observée, l'utilisation à tort d'une moyenne mobile robuste donnera des
résultats très proches des moyennes mobiles classiques et ne dégrade
donc pas les résultats. En revanche, lors de l'estimation en temps réel,
le choix d'une mauvaise spécification (par exemple choc ponctuel,
affecté à l'irrégulier, plutôt que choc permanent, affecté à la
tendance-cycle) peut entraîner des révisions importantes lorsque la
bonne spécification est finalement retenue. Pour minimiser les
révisions, en cas de doute sur la nature du choc il est donc préférable
d'utiliser les moyennes mobiles linéaires classiques avant d'être sûr de
la spécification à adopter.

Afin d'éviter le biais lié à la présence de points atypiques, le
logiciel de désaisonnalisation X-13ARIMA-SEATS (où sont notamment
utilisées les moyennes mobiles de Henderson et de Musgrave) possède
plusieurs modules de correction des chocs : la série est d'abord
pré-corrigée et les chocs sont à la fin réintroduit dans la bonne
composante (saisonnalité, tendance-cycle ou irrégulier). Toutefois, les
rares instituts qui publient des estimations de la tendance-cycle (comme
Statistique Canada et l'Australian Bureau of Statistics) ne semblent pas
utiliser le logiciel X-13ARIMA-SEATS pour l'estimation de cette
composante et appliquent plutôt une moyenne mobile sur le série
désaisonnalisées. Cela peut par exemple s'expliquer lorsque ce ne sont
pas les mêmes équipes qui effectuent la désaisonnalisation et
l'estimation de la tendance-cycle ou pour simplifier le processus. Même
si ces modules de corrections peuvent être appliqués sur les séries
désaisonnalisées (par exemple via un modèle RegARIMA), l'estimation de
l'ampleur du choc dépend du modèle utilisé et de la période d'estimation
retenue, ce qui augmente le risque de mauvaise spécification et
l'incertitude autour des estimations. À l'inverse, l'approche proposée
dans cet article garde les mêmes spécifications que celle des moyennes
mobiles de Henderson et de Musgrave et, par construction, le nombre de
périodes révisées ne dépend que de la longueur de la moyenne mobile.

Dans cette étude, les performances des moyennes mobiles linéaires sont
également comparées à des méthodes non-linéaires robustes pour la
plupart basées sur la médiane. Lorsque les tendances sont lisses et le
bruit faible (par exemple sur des séries simulées), ces méthodes
fournissent des bonnes alternatives aux méthodes classiques en étant
robustes aux chocs ponctuels et en reproduisant, dans la tendance-cycle
les chocs permanents (mais pour qu'ils soient reproduits à la bonne date
il faut attendre l'estimation finale). Toutefois, sur des séries
réelles, les révisions sont plus importantes (avec parfois des
estimations intermédiaires non cohérentes), ce qui suggère, comme
notamment indiqué par \textcite{gather2006online}, d'utiliser des
quantiles plus élevés que la médiane pour les estimations en temps réel.
Par ailleurs, lorsque les points de retournement sont associés à des
chocs (comme lors du COVID-19), en étant naturellement robustes aux
choss, ces méthodes ne vont pas permettre d'estimer une tendance-cycle
qui reproduirait les points de retournement.

En somme, pour l'analyse de la conjoncture et des points de
retournement, l'utilisation de méthodes linéaires locales (telles que
des moyennes mobiles) est à privilégier. Afin de mieux représenter la
réalité économique et de minimiser les révisions, il est important
d'avoir une analyse locale par série : en analysant les chocs et en les
prenant en compte dans la construction des moyennes mobiles comme
proposé dans cette étude ; ou encore en fixant localement les
hyperparamètres utilisés dans les moyennes mobiles asymétriques (lorsque
l'on ne possède pas assez de point dans le futur pour utiliser une
moyenne mobile symétrique), comme suggéré par \textcite{jos2024AQLT},
pour réduire les révisions en évitant des biais systématiques lorsque la
tendance est localement linéaire (par exemple reprise d'activité
progressive après un choc). Toutefois, pour la production de données, il
peut être difficile d'avoir le temps d'analyser chaque série pour
déterminer les chocs. Il peut alors être intéressant d'étudier des
procédures automatiques pour identifier ces chocs, par exemple via des
modèles \emph{ad hoc} (du type modèles RegARIMA) ou bien en reproduisant
le processus de lissage des moyennes mobiles par une modélisation plus
générale (par exemple modélisation espace-état).

\newpage
\appendix

\section{Annexes}\label{sec-annexes}

\subsection{Moyennes mobiles utilisées}\label{sec-annexes-mm}

\begin{figure}[H]

\caption{\label{fig-mm-musgrave}Moyennes mobiles de Henderson et de
Musgrave (\(R=3,5\))}

\centering{

\includegraphics{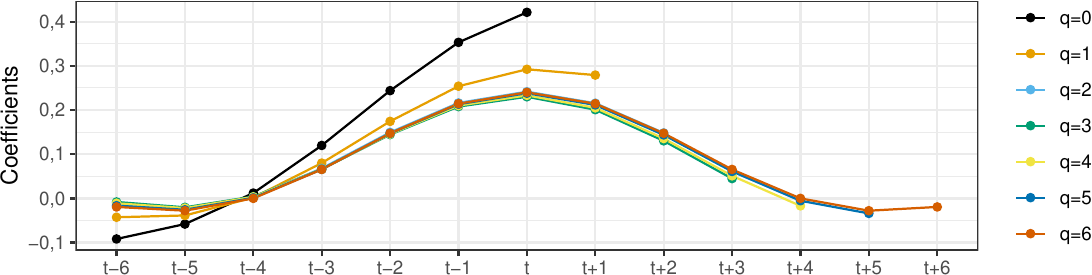}

}

\end{figure}%

\begin{figure}[H]

\caption{\label{fig-clf}Cascade linear filter (CLF) et \emph{Asymmetric
Linear Filter} (ALF) ou méthode « couper-et-normaliser »}

\centering{

\includegraphics{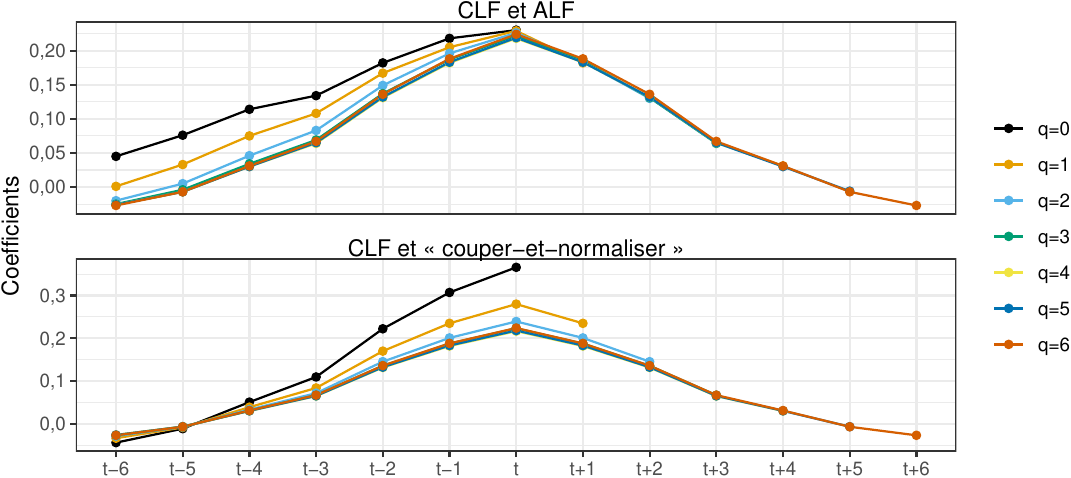}

}

\end{figure}%

\begin{figure}[H]

\caption{\label{fig-robust-ao}Moyennes mobiles de Henderson et de
Musgrave (\(R=3,5\)) robustes à la présence d'un choc ponctuel (AO) à la
dernière date}

\centering{

\includegraphics{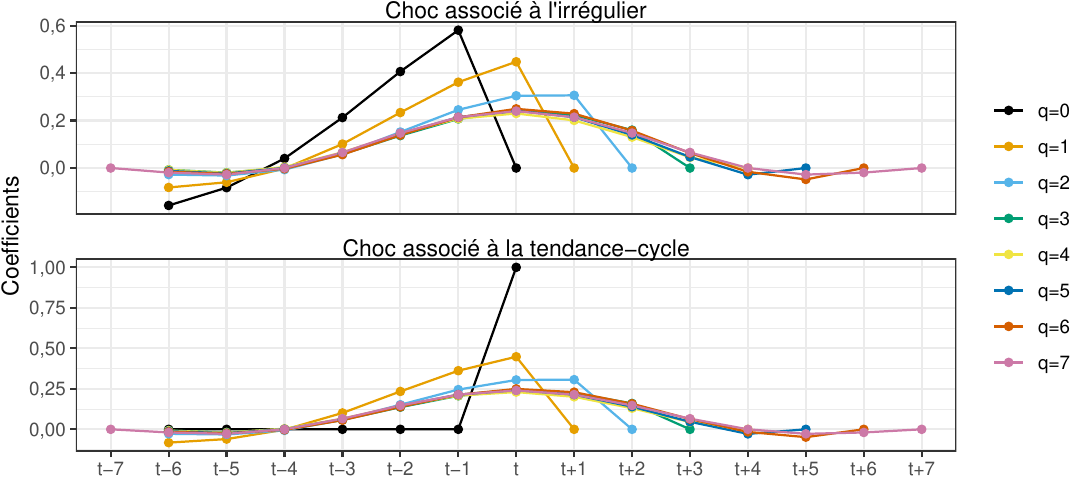}

}

\end{figure}%

\begin{figure}[H]

\caption{\label{fig-robust-aols}Moyennes mobiles de Henderson et de
Musgrave (\(R=3,5\)) à un choc ponctuel (AO) suivi d'un choc permanent
(LS) à la dernière date}

\centering{

\includegraphics{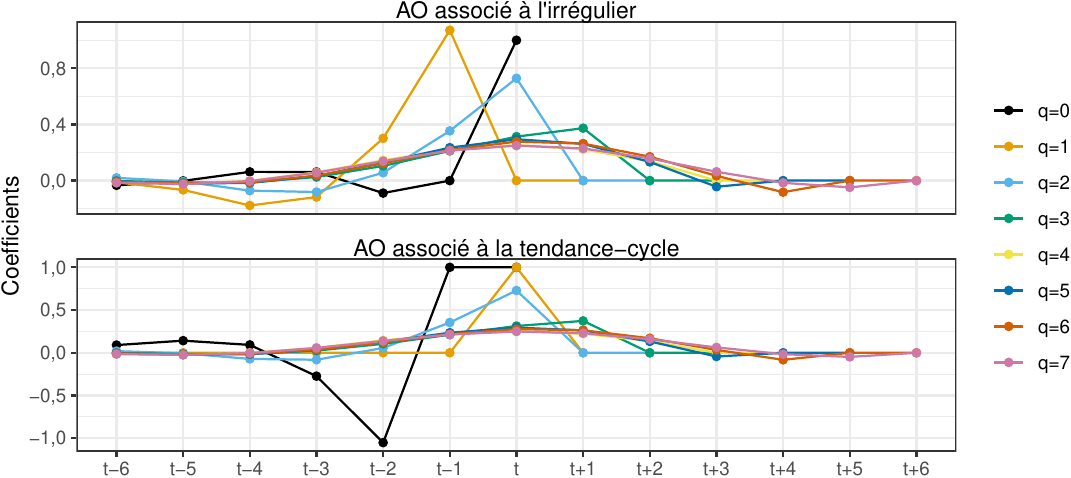}

}

\end{figure}%

\begin{figure}[H]

\caption{\label{fig-robust-ls}Moyennes mobiles de Henderson et de
Musgrave d'un ou deux chocs permanents (LS) présents à la dernière date
(ou deux dernières dates)}

\centering{

\includegraphics{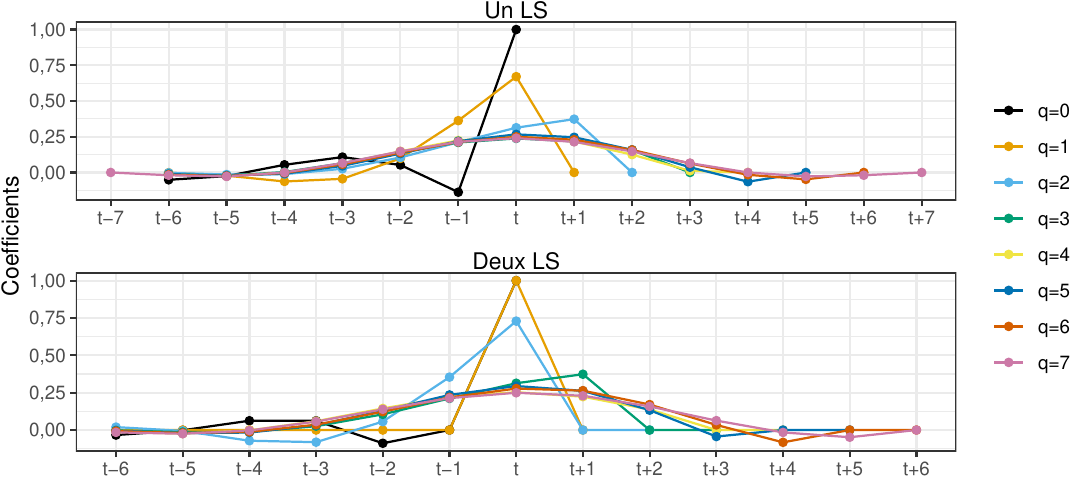}

}

\end{figure}%

\subsection{Calcul du degré de liberté de la variance}\label{sec-df-var}

Soient
\(\boldsymbol\theta = \begin{pmatrix}\theta_{-p},\dots,\theta_f\end{pmatrix}\)
et
\(\boldsymbol \Delta = \transp{(\boldsymbol I - \boldsymbol H)}(\boldsymbol I - \boldsymbol H\boldsymbol)\)
avec : \[
\boldsymbol H=\begin{pmatrix}
&&\boldsymbol0_{p\times n} \\
\theta_{-p} & \cdots & \theta_f  & 0 & \cdots\\
0 & \theta_{-p} & \cdots & \theta_f  & 0 & \cdots
\\ & \ddots &&&\ddots\\
0 &\cdots&0& \theta_{-p} & \cdots & \theta_f \\
&&\boldsymbol0_{f\times n}
\end{pmatrix}\text{ et }
\boldsymbol I = \begin{pmatrix}
&\boldsymbol0_{p\times n} \\
\boldsymbol 0_{(n-p-f)\times p} &\boldsymbol I_{n-p-f} & \boldsymbol 0_{(n-p-f)\times f}  \\
&\boldsymbol0_{f\times n}
\end{pmatrix},
\] où \(\boldsymbol 0_{p\times n}\) est la matrice de taille
\(p\times n\) ne contenant que de zéros et \(\boldsymbol I_{n-p-f}\) la
matrice identité de taille \(n-p-f.\)

L'objectif de cette annexe est de montrer que le calcul de
\(\tr(\boldsymbol \Delta^2)\) peut se simplifier par le produit d'une
matrice de taille \(1\times (p+f+1)\) et d'une matrice de taille
\((p+f+1)\times(p+f+1)\), ce qui permet de réduire le temps de calcul du
degré de liberté de la loi de Student utilisée pour la construction
d'intervalles de confiance : \[
\tr({\boldsymbol \Delta}^2)=(n-(p+f))L_0^2+2\sum_{k=1}^{p+f}(n-(p+f)-k)L_k^2
\] où \(L_0,\dots,L_{p+f}\) sont définis par : \[
\begin{pmatrix}w_{-p} & \cdots & w_f\end{pmatrix}
\begin{pmatrix}
w_{-p} &0 & 0 &\cdots& 0 \\
w_{-p+1} & w_{-p} & 0 & \ddots & \vdots \\
w_{-p+2} & w_{-p+1}& w_{-p} & \ddots & \vdots\\
\vdots & \vdots & \vdots & \ddots &\vdots \\
w_f & w_{f-1} & w_{f-2}&\ddots&w_{-p}\\
\end{pmatrix}=\begin{pmatrix} L_0 & \cdots & L_{p+f} \end{pmatrix}
\] avec \(\boldsymbol w = \begin{pmatrix}w_{-p},\dots,w_f\end{pmatrix}\)
la moyenne mobile telle que \(w_i=\1_{i=0}-\theta_i\).

Soient \(i,j\in\{1,\dots,n\},\) posons par convention \(w_i=0\) pour
\(i\notin[-p,f].\) Notons
\(\boldsymbol \Gamma=\boldsymbol I -\boldsymbol H.\) On a
\({\boldsymbol H}_{i,j}=\theta_{i-j}\1_{i\in[p+1,n-f]}\) et donc
\({\boldsymbol \Gamma}_{i,j}=w_{i-j}\1_{i\in[p+1,n-f]}\) et ensuite :
\begin{align*}
(\transp{{\boldsymbol \Gamma}}{\boldsymbol \Gamma})_{i,j} &=\sum_{k=1}^n{\boldsymbol \Gamma}_{k,i}{\boldsymbol \Gamma}_{k,j}=\sum_{k=1}^nw_{k-i}w_{k-j} \1_{(i,j)\in[p+1,n-f]^2}\\
&=\sum_{k=-p}^fw_{k}w_{i-j+k}\1_{(i,j)\in[p+1,n-f]^2}.
\end{align*} Puis : \begin{align*}
((\transp{{\boldsymbol \Gamma}}{\boldsymbol \Gamma})^2)_{i,i} &= \sum_{j=1}^n(\transp{{\boldsymbol \Gamma}}{\boldsymbol \Gamma})_{i,j}(\transp{{\boldsymbol \Gamma}}{\boldsymbol \Gamma})_{j,i}\\
&=\sum_{j=1}^n(\transp{{\boldsymbol \Gamma}}{\boldsymbol \Gamma})_{i,j}^2\text{ car la matrice }\transp{{\boldsymbol \Gamma}}{\boldsymbol \Gamma}\text{ est symétrique} \\
&=\sum_{j=1}^n\left(\sum_{k=-p}^fw_{k}w_{i-j/k}\1_{(i,j)\in[p+1,n-f]^2}\right)^2\\
&=\sum_{j=p}^{n-f}\left(\sum_{k=-p}^fw_{k}w_{i-j+k}\right)^2\1_{i\in[p+1,n-f]}.
\end{align*} Donc
\(((\transp{{\boldsymbol \Gamma}}{\boldsymbol \Gamma})^2)_{i,i} = 0\) si
\(i\notin [p+1,n-f].\) Pour \(i\in[p+1,n-f],\) comme pour tout
\(k\in [-p,f]\) on a \(w_{i-j+k} =0\) si \(j<i-(f+p)\) ou si
\(j>i-(f+p)\), il vient : \begin{align*}
((\transp{{\boldsymbol \Gamma}}{\boldsymbol \Gamma})^2)_{i,i} &= \sum_{j=\max(i-(f+p),1)}^{\min(i+(f+p),n)}\left(\sum_{k=-p}^fw_{k}w_{i-j+k}\right)^2 \\
&=\sum_{j=-(f+p)}^{f+p}\left(\sum_{k=-p}^fw_{k}w_{k-j}\right)^2\1_{j\in [1-i,n-i]} \\
&=\sum_{j=-(f+p)}^{f+p}\left(L_{|j|}\right)^2\1_{j\in [1-i,n-i]}\text{ par symétrie autour de }0.
\end{align*} En somme : \begin{align*}
\tr({\boldsymbol \Delta^2})& = \sum_{i=p}^{n-f} \sum_{j=-(f+p)}^{f+p}\left(L_{|j|}\right)^2\1_{j\in [1-i,n-i]} \\
&= (n-(p+f))L_0^2+2(n-(p+f)-1)L_1^2 + 2(n-(p+f)-2)L_2^2 +\\
&\phantom{=}\dots+2(n-2(p+f))L_{p+f}^2 \\
&=(n-(p+f))L_0^2+2\sum_{k=1}^{p+f}(n-(p+f)-k)L_k^2.
\end{align*}

Cette optimisation computationnelle a ensuite été implémentée dans la
fonction \texttt{rjd3filters::confint\_filter()}.

\subsection{Autres exemples sur des séries réelles}\label{sec-autres-ex}

\subsubsection{\texorpdfstring{Chocs ponctuels (\emph{additive outlier},
AO)}{Chocs ponctuels (additive outlier, AO)}}\label{sec-autres-ex-ao}

\begin{figure}[H]

\caption{\label{fig-ipi-voitures-out-y}IPI CVS-CJO dans la construction
de véhicules automobiles autour des points atypiques d'août 1998, 1999
et 2004}

\centering{

\includegraphics{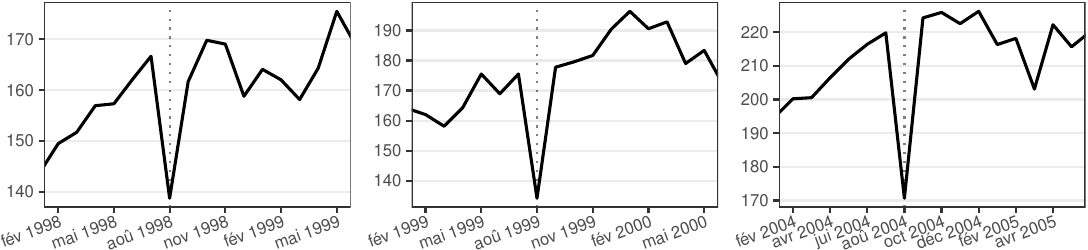}

}

\end{figure}

Les chocs ponctuels (\emph{additive outlier}, AO) sont illustrés sur
l'indice de la production industrielle (IPI) dans la construction de
véhicules automobiles, corrigé des variations saisonnières (CVS) et des
jours ouvrables (CJO), publié par l'Insee (série
\href{https://www.insee.fr/fr/statistiques/serie/010768140}{\texttt{010768140}}
publiée le 04 octobre 2024). Trois chocs ponctuels s'observent
clairement sur les mois d'août de 1998, 1999 et 2004
(figure~\ref{fig-ipi-voitures-out-y}).

\begin{figure}[H]

\caption{\label{fig-ipi-voitures98-out1-est}Estimations en temps réel de
la tendance-cycle pour l'IPI dans la construction de véhicules
automobiles à partir d'août 1998}

\centering{

\includegraphics{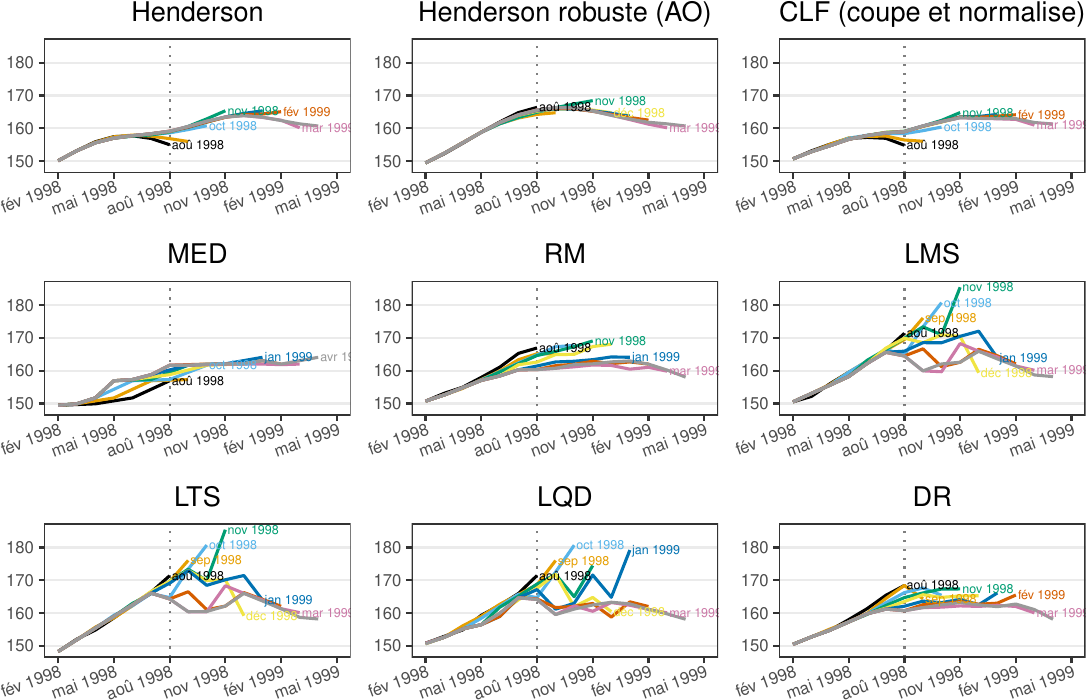}

\scriptsize\raggedright

\textbf{Note} : Henderson : Henderson (estimation finale) et Musgrave
(estimations intermédiaires) ; Henderson robuste : moyennes mobiles
robustes de Henderson (estimation finale) et de Musgrave (estimations
intermédiaires) ; CLF : \emph{cascade linear filter} (estimation finale)
et méthode « couper-et-normaliser » (estimations intermédiaires) ; MED :
Médiane mobile ; RM : médiane répétée ; LMS : moindres carrés médians ;
LTS : moindres carrés élagués ; LQD : moindres quartiles différenciés ;
DR : Régression profonde.

}

\end{figure}%

\begin{figure}[H]

\caption{\label{fig-ipi-voitures98-out1-ci}Intervalles de confiance pour
les filtres de Henderson et filtres de Henderson robustes robuste pour
l'IPI dans la construction de véhicules automobiles à partir d'août
1998}

\centering{

\includegraphics{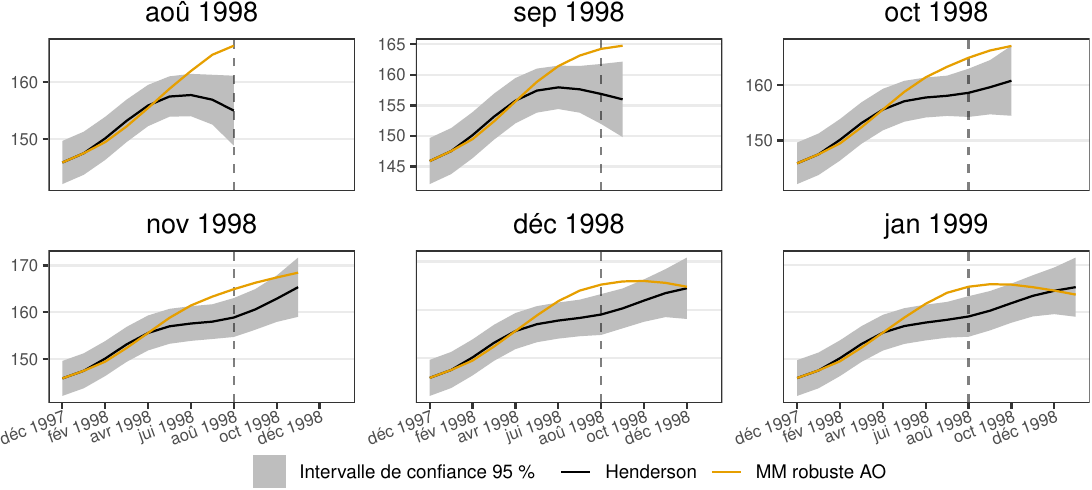}

}

\end{figure}%

\begin{figure}[H]

\caption{\label{fig-ipi-voitures98-out2-est}Estimations en temps réel de
la tendance-cycle pour l'IPI dans la construction de véhicules
automobiles à partir d'août 1999}

\centering{

\includegraphics{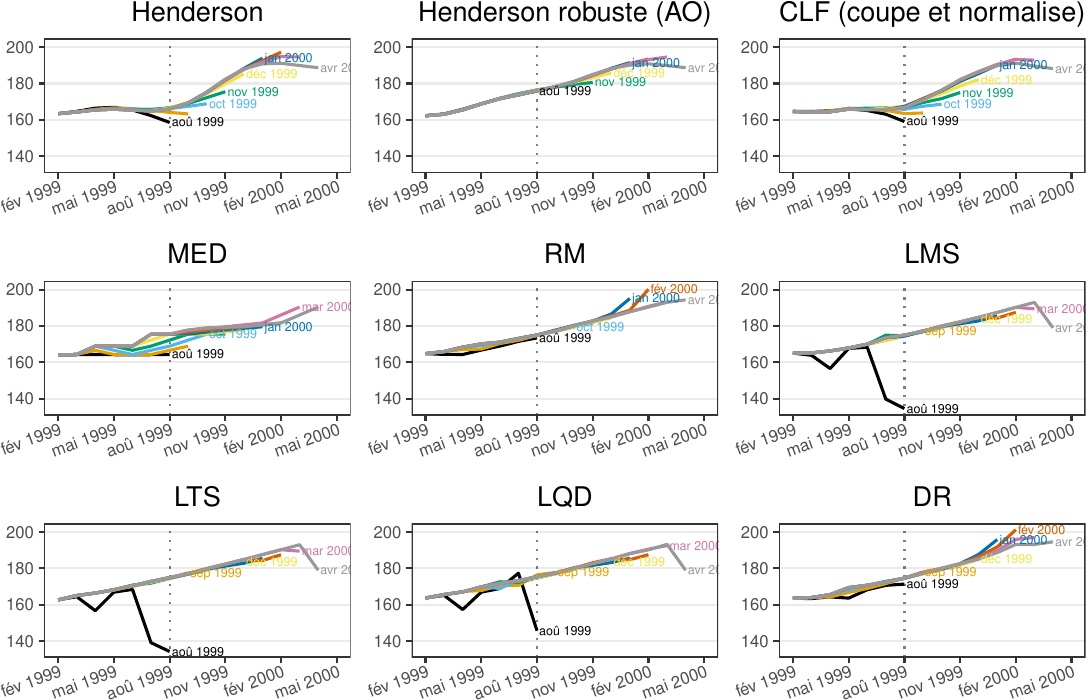}

\scriptsize\raggedright

\textbf{Note} : Henderson : Henderson (estimation finale) et Musgrave
(estimations intermédiaires) ; Henderson robuste : moyennes mobiles
robustes de Henderson (estimation finale) et de Musgrave (estimations
intermédiaires) ; CLF : \emph{cascade linear filter} (estimation finale)
et méthode « couper-et-normaliser » (estimations intermédiaires) ; MED :
Médiane mobile ; RM : médiane répétée ; LMS : moindres carrés médians ;
LTS : moindres carrés élagués ; LQD : moindres quartiles différenciés ;
DR : Régression profonde.

}

\end{figure}%

\begin{figure}[H]

\caption{\label{fig-ipi-voitures98-out2-ci}Intervalles de confiance pour
les filtres de Henderson et filtres de Henderson robustes robuste pour
l'IPI dans la construction de véhicules automobiles à partir d'août
1999}

\centering{

\includegraphics{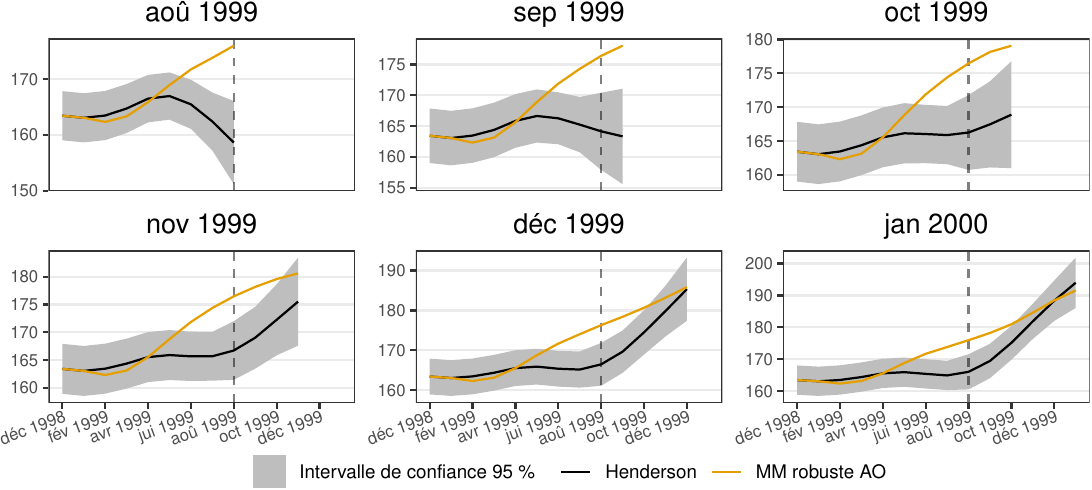}

}

\end{figure}%

\begin{figure}[H]

\caption{\label{fig-ipi-voitures2004-est}Estimations en temps réel de la
tendance-cycle pour l'IPI dans la construction de véhicules automobiles
à partir d'août 2004}

\centering{

\includegraphics{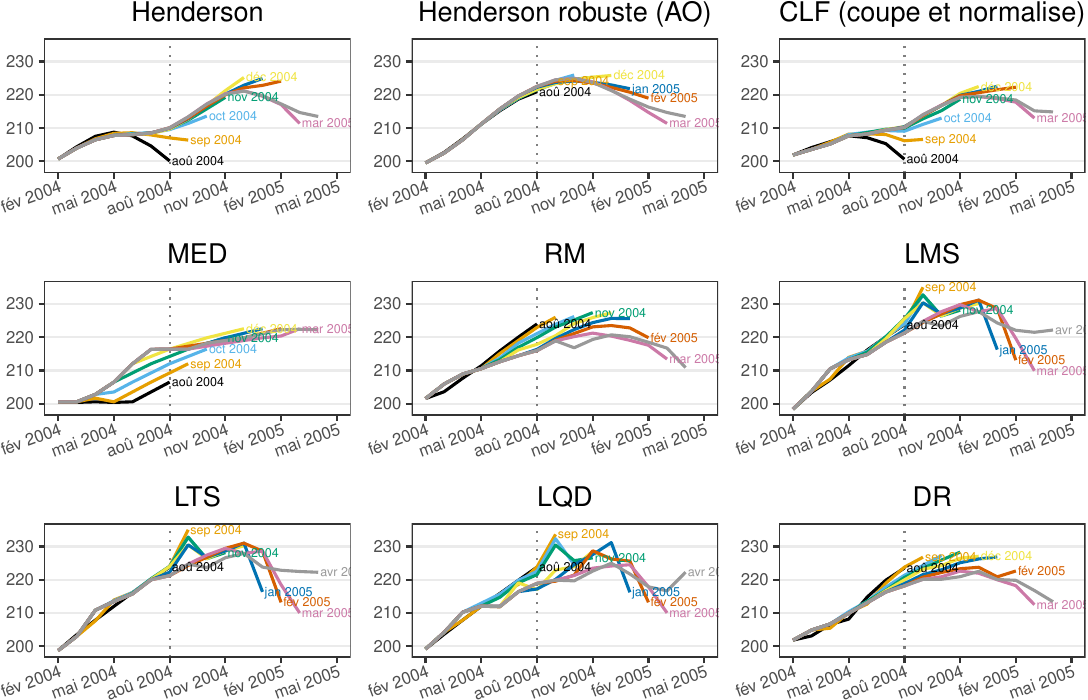}

\scriptsize\raggedright

\textbf{Note} : Henderson : Henderson (estimation finale) et Musgrave
(estimations intermédiaires) ; Henderson robuste : moyennes mobiles
robustes de Henderson (estimation finale) et de Musgrave (estimations
intermédiaires) ; CLF : \emph{cascade linear filter} (estimation finale)
et méthode « couper-et-normaliser » (estimations intermédiaires) ; MED :
Médiane mobile ; RM : médiane répétée ; LMS : moindres carrés médians ;
LTS : moindres carrés élagués ; LQD : moindres quartiles différenciés ;
DR : Régression profonde.

}

\end{figure}%

\begin{figure}[H]

\caption{\label{fig-ipi-voitures2004-ci}Intervalles de confiance pour
les filtres de Henderson et filtres de Henderson robustes robuste pour
l'IPI dans la construction de véhicules automobiles à partir d'août
2004}

\centering{

\includegraphics{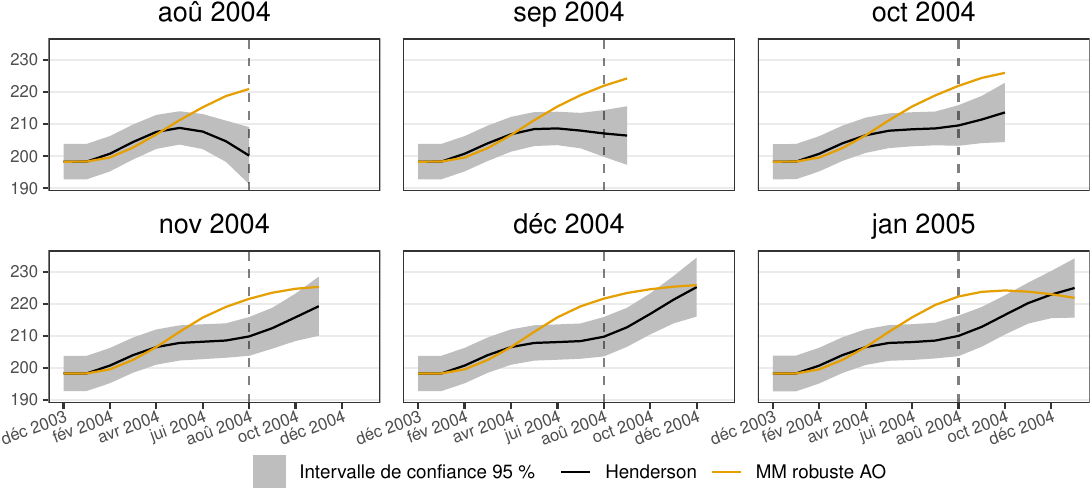}

}

\end{figure}%

\subsubsection{\texorpdfstring{Choc permanent (\emph{level shift},
LS)}{Choc permanent (level shift, LS)}}\label{sec-autres-ex-ls}

Les chocs permanents (\emph{level shift}, LS) sont illustrés sur
l'indice de la production industrielle (IPI) dans l'extraction de
pétrole brut, corrigé des variations saisonnières (CVS) et des jours
ouvrables (CJO), publié par l'Insee (série
\href{https://www.insee.fr/fr/statistiques/serie/010767578}{\texttt{010767578}})
le 04 octobre 2024 (données alors disponibles jusqu'en août 2024). Un
choc permanent s'observe en juin 2010. Pour la construction des
intervalles de confiance, la série est tronquée pour ne prendre que les
valeurs après janvier 2001 avant d'éviter que le début de la série, dont
le profil est assez différent de celui sur la fin de la série,
n'augmente l'estimation de la variance de l'irrégulier.

\begin{figure}[H]

\caption{\label{fig-ipi-petrole-brut10-y}IPI CVS-CJO dans l'extraction
de pétrole brut autour du choc permanent de juin 2010}

\centering{

\includegraphics{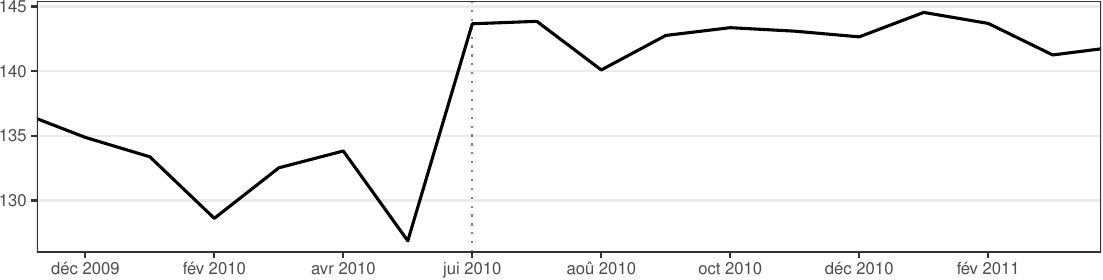}

}

\end{figure}%

\begin{figure}[H]

\caption{\label{fig-ipi-petrole-brut10-est}Estimations en temps réel de
la tendance-cycle pour l'IPI dans l'extraction de pétrole brut à partir
de juin 2010}

\centering{

\includegraphics{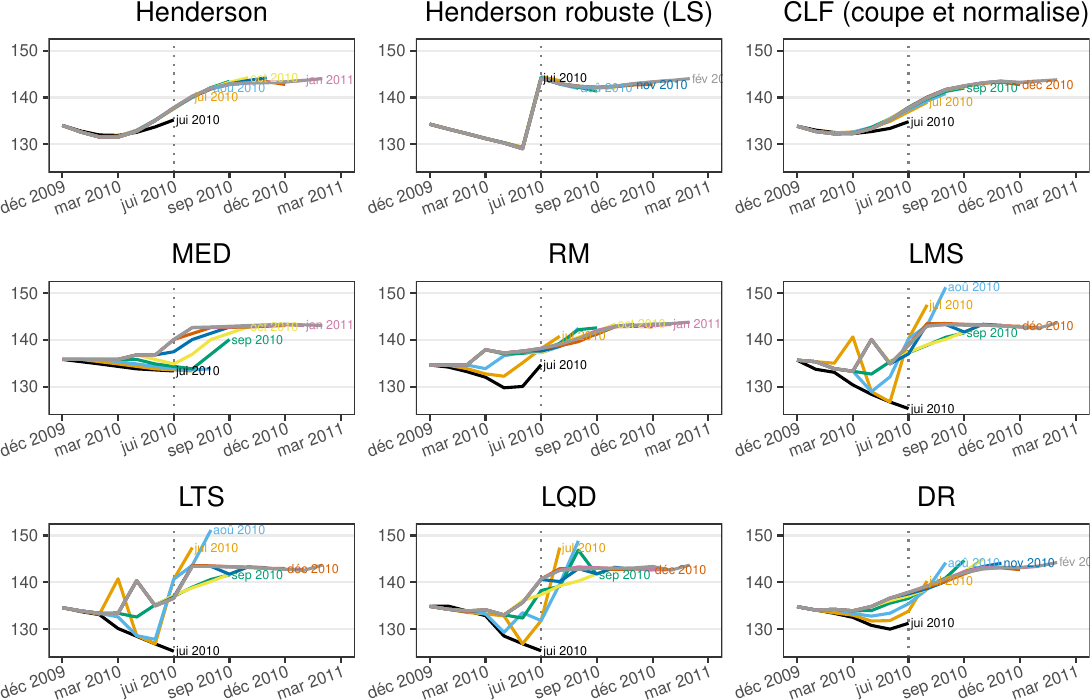}

\scriptsize\raggedright

\textbf{Note} : Henderson : Henderson (estimation finale) et Musgrave
(estimations intermédiaires) ; Henderson robuste : moyennes mobiles
robustes de Henderson (estimation finale) et de Musgrave (estimations
intermédiaires) ; CLF : \emph{cascade linear filter} (estimation finale)
et méthode « couper-et-normaliser » (estimations intermédiaires) ; MED :
Médiane mobile ; RM : médiane répétée ; LMS : moindres carrés médians ;
LTS : moindres carrés élagués ; LQD : moindres quartiles différenciés ;
DR : Régression profonde.

}

\end{figure}%

\begin{figure}[H]

\caption{\label{fig-ipi-petrole-brut10-ci}Intervalles de confiance pour
les filtres de Henderson et filtres de Henderson robustes robuste pour
l'IPI dans l'extraction de pétrole brut à partir de juin 2010}

\centering{

\includegraphics{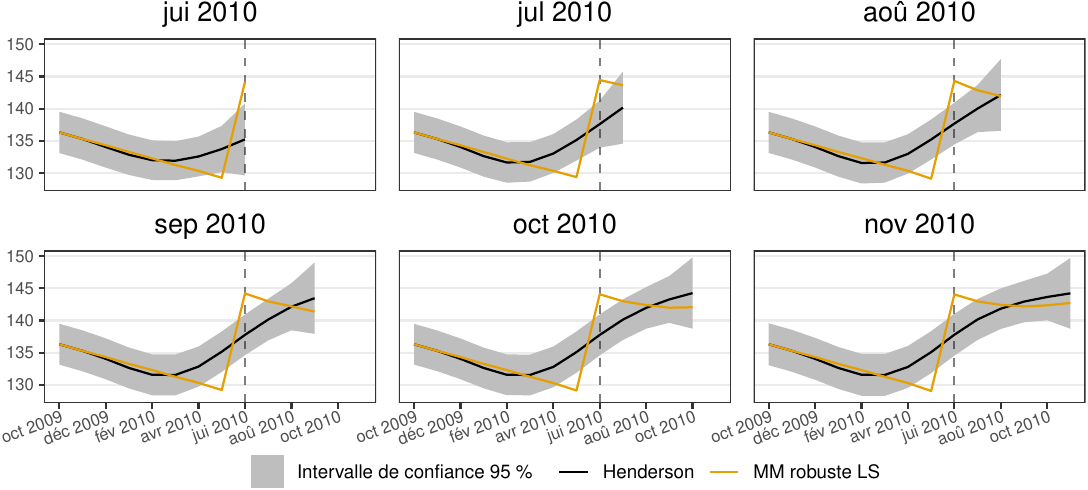}

}

\end{figure}%

\subsubsection{COVID-19}\label{sec-autres-ex-covid}

\begin{figure}[H]

\caption{\label{fig-autresex-covid-y}Emploi et ventes au détail et
services de restauration aux États-Unis autour de mars 2020}

\centering{

\includegraphics{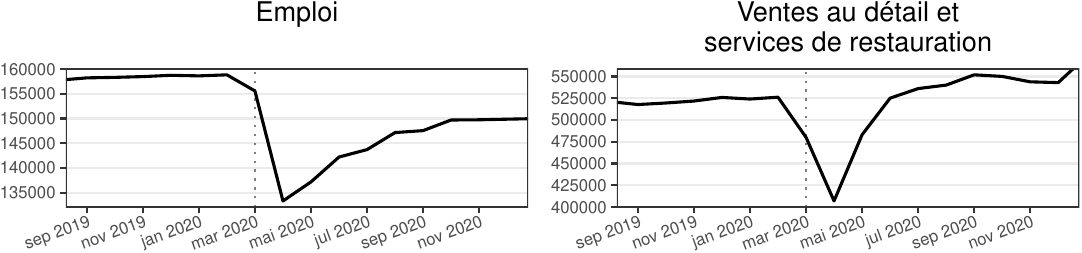}

}

\end{figure}

Deux autres exemples de l'estimation de la tendance-cycle pendant le
COVID-19 sont données à partir du niveau d'emploi (série
\texttt{CE16OV}) et des ventes au détail et services de restauration
(série `RETAILx) aux États-Unis, séries issues de la base FRED-MD. Les
points de retournement officiels de la datation du National Bureau of
Economic Research (NBER) sont en février et avril 2020\footnote{
  \url{https://www.nber.org/research/data/us-business-cycle-expansions-and-contractions}}.
Pour l'emploi le retour à la normal est progressif à partir d'avril 2020
alors qu'il est rapide pour les ventes au détail et services de
restauration (figure~\ref{fig-autresex-covid-y}).

\begin{figure}[H]

\caption{\label{fig-ce16ov-covid-est}Emploi aux États-Unis à partir de
mars 2020}

\centering{

\includegraphics{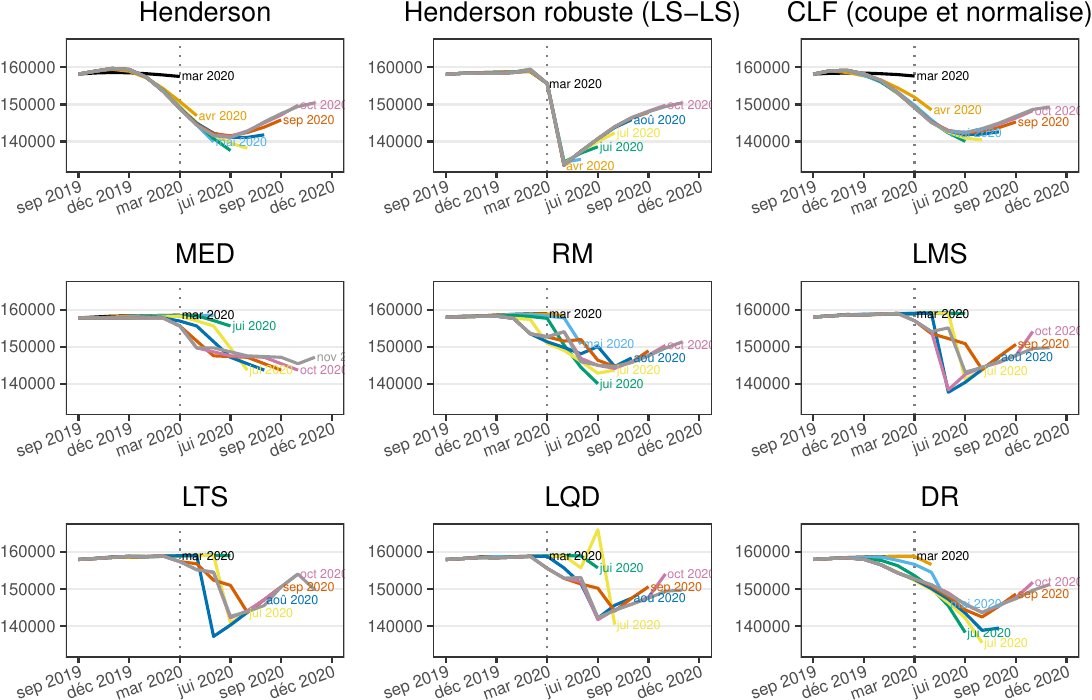}

}

\end{figure}

\begin{figure}[H]

\caption{\label{fig-ce16ov-covid-ci}Intervalles de confiance pour les
filtres de Henderson et filtres de Henderson robustes robuste pour
l'emploi aux États-Unis à partir de mars 2020}

\centering{

\includegraphics{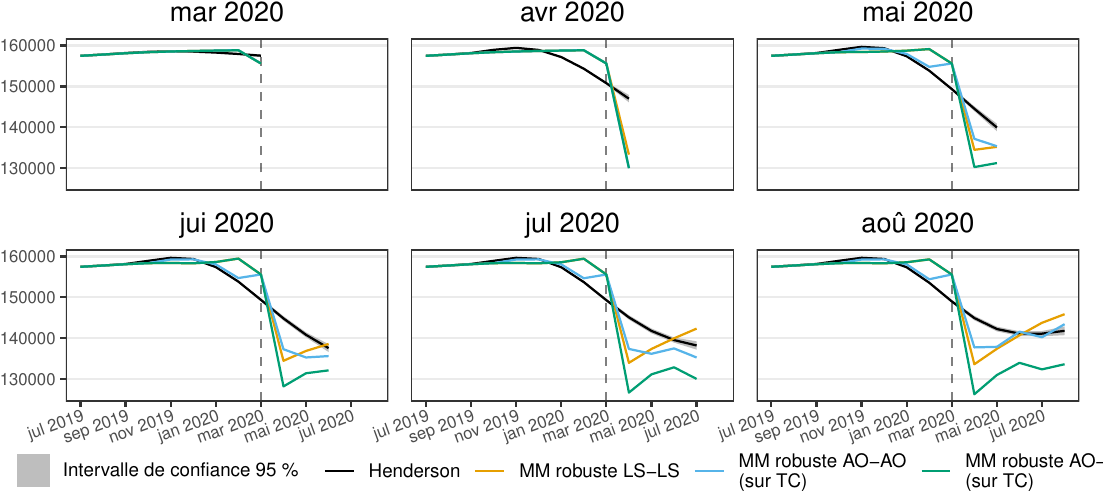}

}

\end{figure}

\begin{figure}[H]

\caption{\label{fig-retailx-covid-est}Estimations en temps réel de la
tendance-cycle des ventes au détail et services de restauration aux
États-Unis à partir de mars 2020}

\centering{

\includegraphics{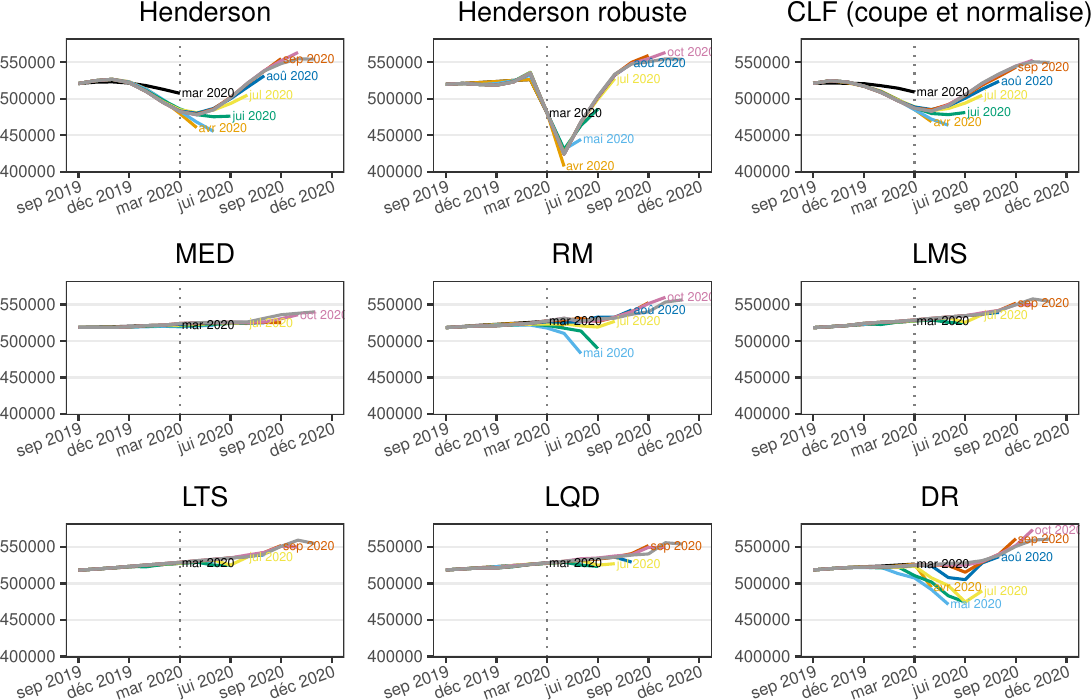}

\scriptsize\raggedright

\textbf{Note} : Henderson : Henderson (estimation finale) et Musgrave
(estimations intermédiaires) ; Henderson robuste : moyennes mobiles
robustes de Henderson (estimation finale) et de Musgrave (estimations
intermédiaires) ; CLF : \emph{cascade linear filter} (estimation finale)
et méthode « couper-et-normaliser » (estimations intermédiaires) ; MED :
Médiane mobile ; RM : médiane répétée ; LMS : moindres carrés médians ;
LTS : moindres carrés élagués ; LQD : moindres quartiles différenciés ;
DR : Régression profonde.

}

\end{figure}

\begin{figure}[H]

\caption{\label{fig-retailx-covid-ci}Intervalles de confiance pour les
filtres de Henderson et filtres de Henderson robustes robuste pour les
ventes au détail et services de restauration aux États-Unis à partir de
mars 2020}

\centering{

\includegraphics{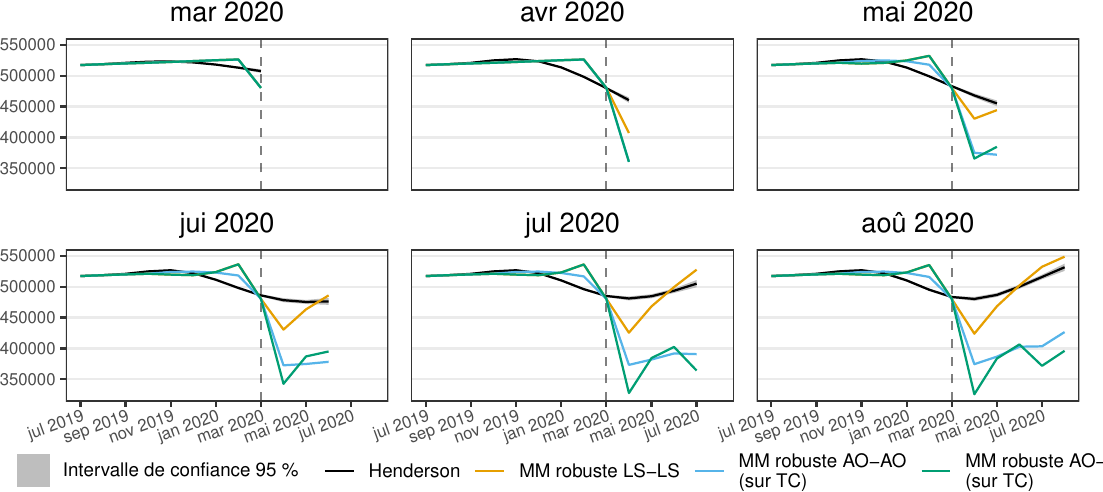}

}

\end{figure}

Comme pour l'IPI dans l'industrie manufacturière, les méthodes robustes
effacent le choc (figures \ref{fig-ce16ov-covid-est} et
\ref{fig-retailx-covid-est}), ce qui ne permet pas d'analyser les points
de retournement. Les moyennes mobiles linéaires classiques
(Henderson/Musgrave et CLF) détectent les points de retournement avec un
décalage : pic en octobre 2019 et creux en avril 2020 pour les ventes au
détail et services de restauration et juin 2020 pour l'emploi. Pour les
moyennes mobiles robustes, le choc est modélisé par deux chocs
permanents en mars et avril 2020. Les révisions sont faibles et cela
permet de bien détecter les deux retournement conjoncturels
contrairement aux moyennes mobiles robustes où l'on considère mars et
avril 2020 comme des chocs ponctuels et associant le choc à la
tendance-cycle (figures \ref{fig-ce16ov-covid-ci} et
\ref{fig-retailx-covid-ci}).

\subsubsection{Points de retournement}\label{sec-autres-ex-retournement}

Un autre exemple de point de retournement est illustré à partir des
ventes au détail et services de restauration aux États-Unis (série
\texttt{RETAILx} de la base FRED-MD) autour du point de retournement de
novembre 2007.

\begin{figure}[H]

\caption{\label{fig-retailx2007-y}Ventes au détail et services de
restauration aux États-Unis autour du point de retournement de novembre
2007}

\centering{

\includegraphics{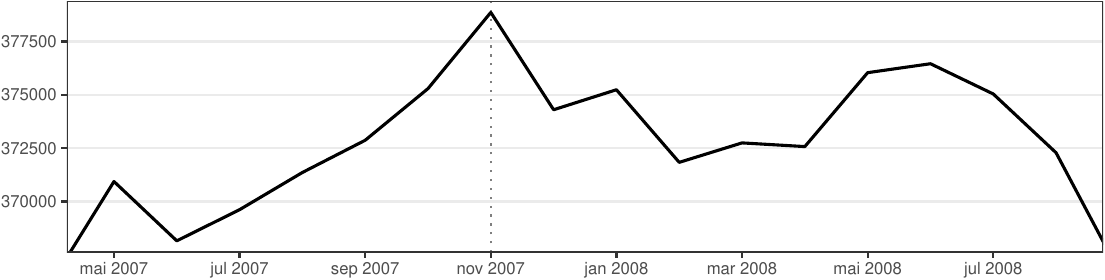}

}

\end{figure}%

\begin{figure}[H]

\caption{\label{fig-retailx2007-est}Estimations en temps réel de la
tendance-cycle des ventes au détail et services de restauration aux
États-Unis à partir de novembre 2007}

\centering{

\includegraphics{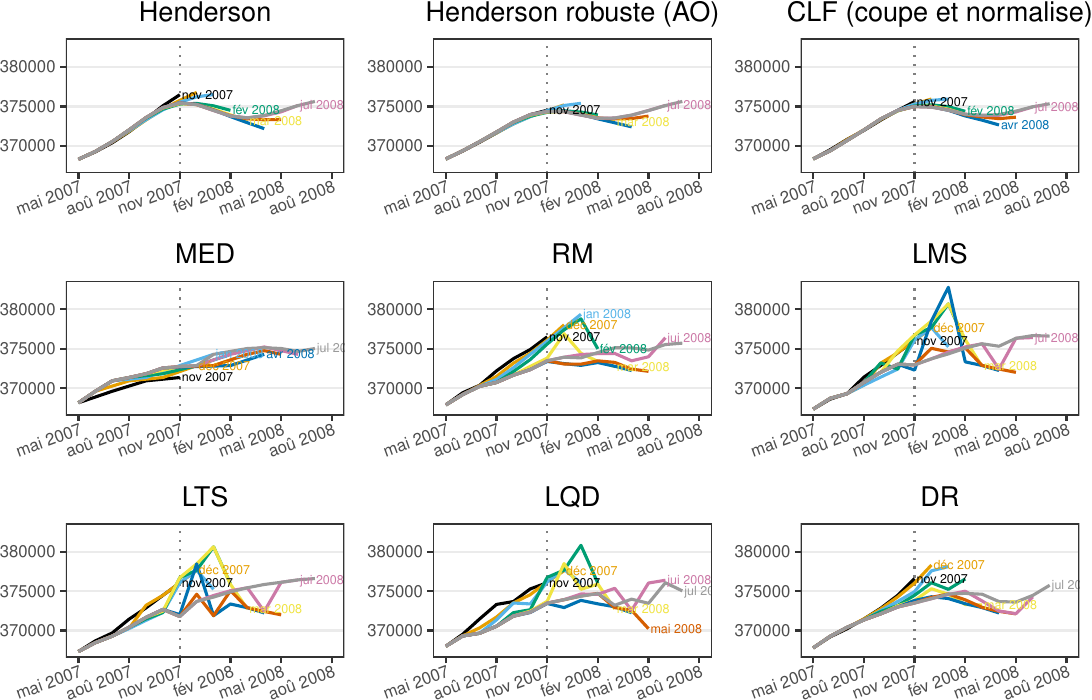}

\scriptsize\raggedright

\textbf{Note} : Henderson : Henderson (estimation finale) et Musgrave
(estimations intermédiaires) ; Henderson robuste : moyennes mobiles
robustes de Henderson (estimation finale) et de Musgrave (estimations
intermédiaires) ; CLF : \emph{cascade linear filter} (estimation finale)
et méthode « couper-et-normaliser » (estimations intermédiaires) ; MED :
Médiane mobile ; RM : médiane répétée ; LMS : moindres carrés médians ;
LTS : moindres carrés élagués ; LQD : moindres quartiles différenciés ;
DR : Régression profonde.

}

\end{figure}%

\begin{figure}[H]

\caption{\label{fig-retailx2007-ci}Intervalles de confiance pour les
filtres de Henderson et filtres de Henderson robustes robuste pour les
ventes au détail et services de restauration aux États-Unis à partir de
novembre 2007}

\centering{

\includegraphics{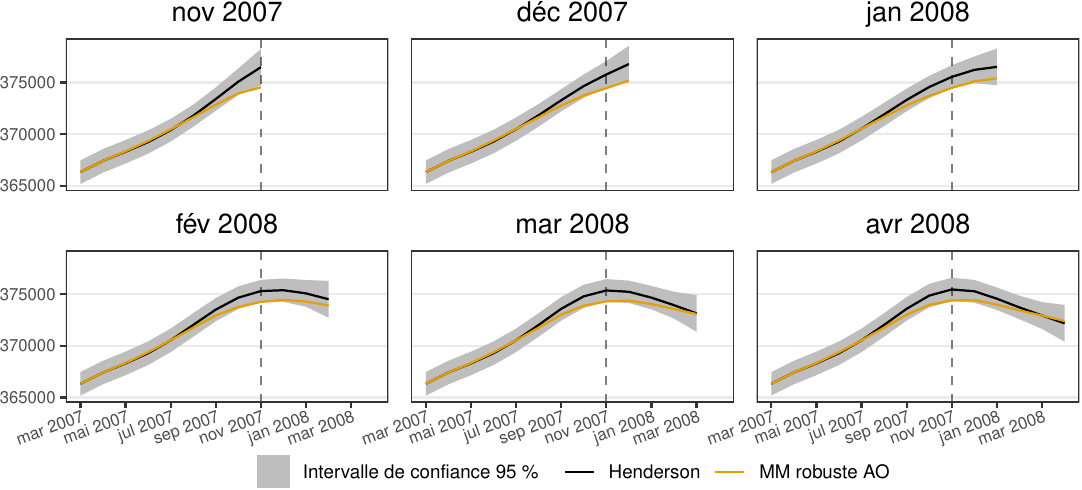}

}

\end{figure}%

\clearpage

\subsection{Moindres carrés médians (LMS) et moindres carrés élagués
(LTS) avec une tendance de degré 2}\label{sec-an-lms-lts}

\begin{figure}[H]

\caption{\label{fig-lmsd2}Estimations en temps réel de la tendance-cycle
pour la méthode des moindres carrés médians (LMS) en modélisant une
tendance locale de degré 2}

\centering{

\includegraphics{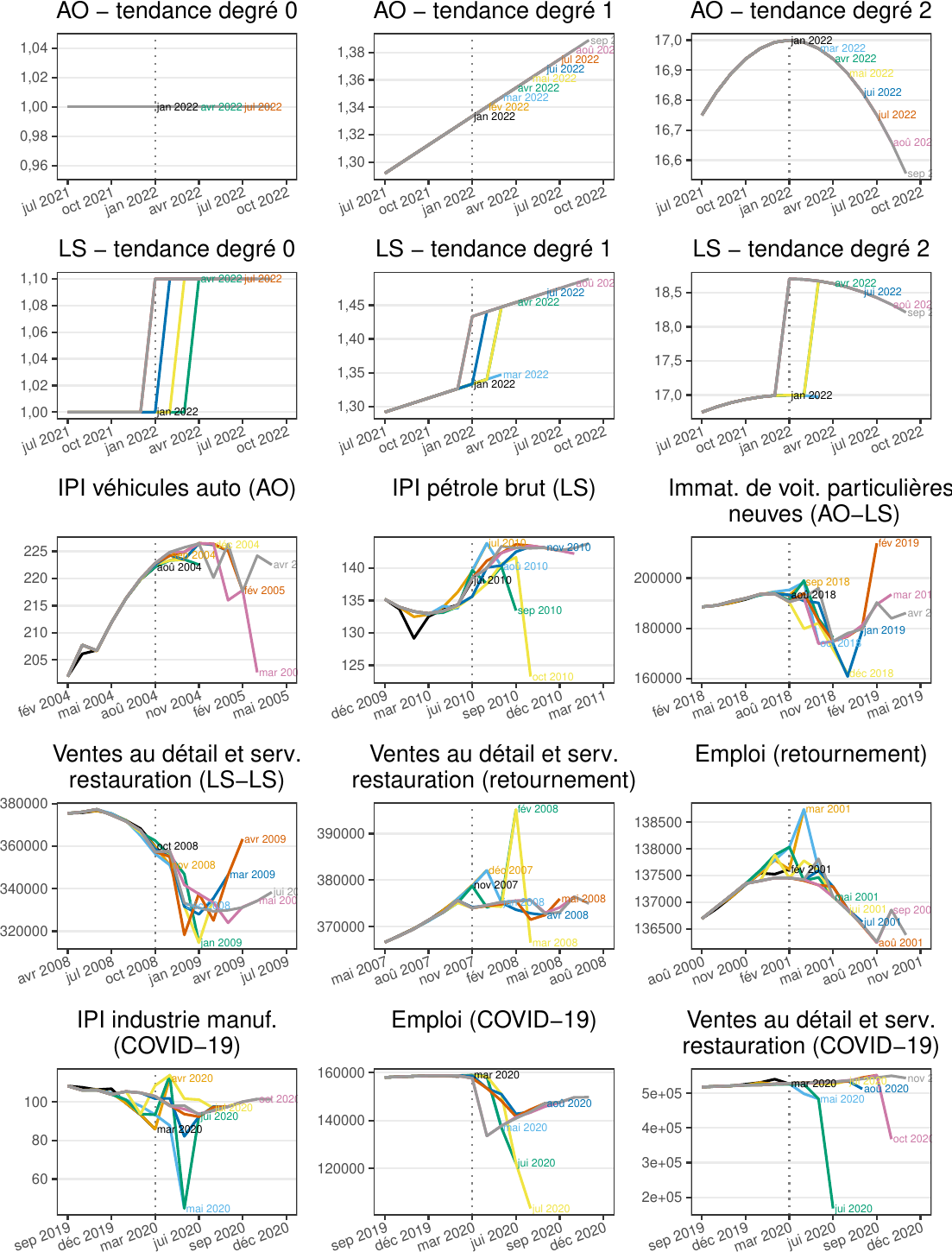}

}

\end{figure}

\begin{figure}[H]

\caption{\label{fig-ltsd2}Estimations en temps réel de la tendance-cycle
pour la méthode des moindres carrés élagués (LTS) en modélisant une
tendance locale de degré 2}

\centering{

\includegraphics{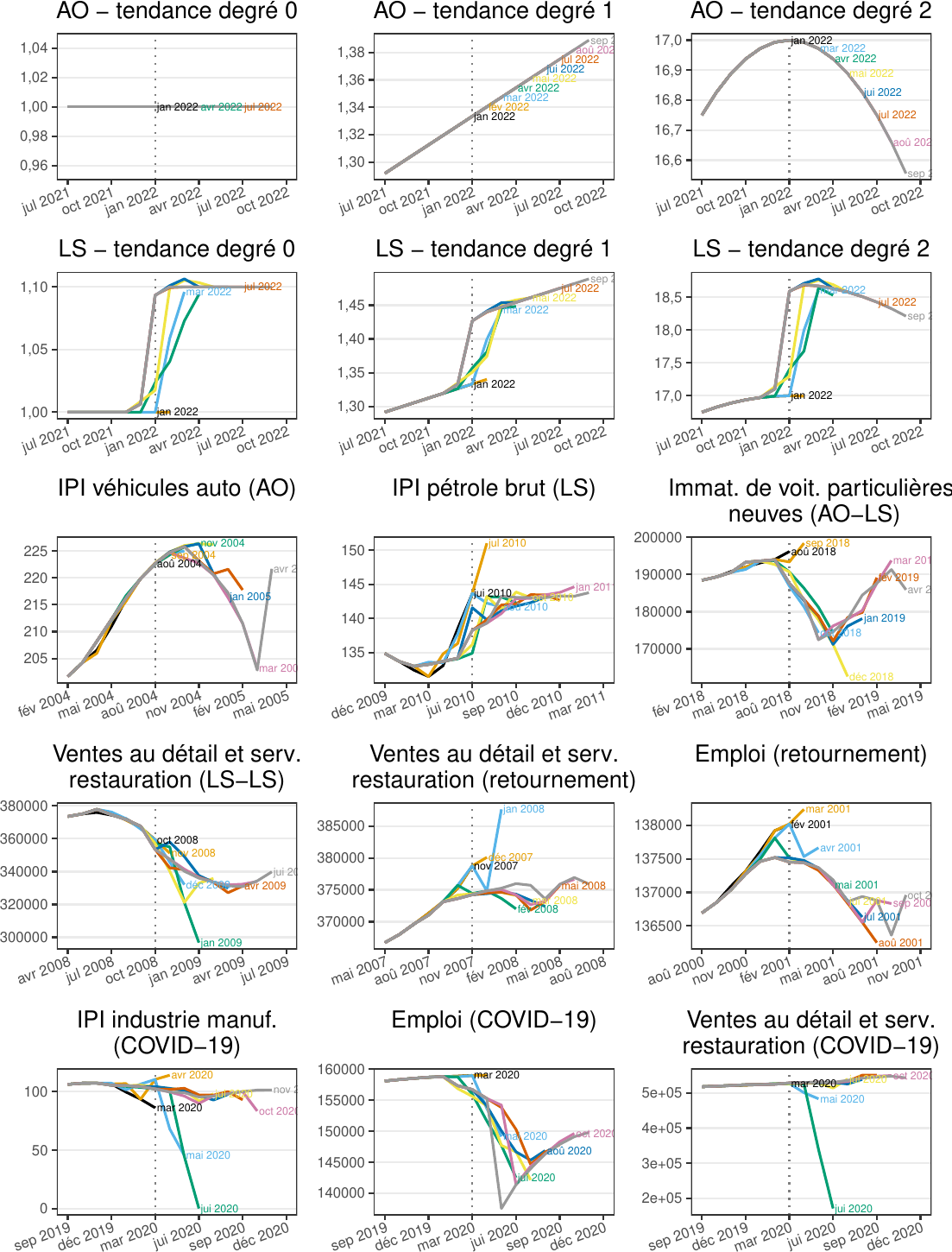}

}

\end{figure}

Cette annexe présente les résultats des estimations de la
réimplémentation des processus d'estimations locale des moindres carrés
médians (LMS, figure~\ref{fig-lmsd2}) et moindres carrés élagués (LTS,
figure~\ref{fig-ltsd2}) avec une tendance de degré 2.

\clearpage

\section*{Bibliographie}\label{bibliographie}
\addcontentsline{toc}{section}{Bibliographie}

\printbibliography[heading=none]

\end{document}